\documentclass[twocolumn]{aastex62}
\usepackage{graphicx, natbib, hyperref}
\usepackage[flushleft]{threeparttable}
\hypersetup{linkcolor=red,citecolor=green,filecolor=cyan,urlcolor=magenta}

\received{xx}
\revised{xx}
\accepted{xx}
\submitjournal{ApJ}

\shorttitle{ASPECS: Molecular gas in CO-selected galaxies} 
\shortauthors{Aravena et al.}

   \def\lsim{\mathrel{\rlap{\lower
3pt \hbox{$\sim$}} \raise 2.0pt \hbox{$<$}}} \def\gsim{\mathrel{\rlap{\lower 3pt \hbox{$\sim$}}
\raise 2.0pt \hbox{$>$}}}

\begin{document}

\title{The ALMA Spectroscopic Survey in the Hubble Ultra Deep Field: Evolution of the molecular gas in CO-selected galaxies}


\author[0000-0002-6290-3198]{Manuel Aravena}\email{manuel.aravenaa@mail.udp.cl}
\affil{N\'{u}cleo de Astronom\'{\i}a, Facultad de Ingenier\'{\i}a y Ciencias, Universidad Diego Portales, Av. Ej\'{e}rcito 441, Santiago, Chile}

\author[0000-0002-2662-8803]{Roberto Decarli}
\affil{INAF?Osservatorio di Astrofisica e Scienza dello Spazio, via Gobetti 93/3, I-40129, Bologna, Italy}

\author{Jorge G\'onzalez-L\'opez}
\affil{N\'{u}cleo de Astronom\'{\i}a, Facultad de Ingenier\'{\i}a y Ciencias, Universidad Diego Portales, Av. Ej\'{e}rcito 441, Santiago, Chile}

\author{Leindert Boogaard}
\affil{Leiden Observatory, Leiden University, PO Box 9513, NL2300 RA Leiden, The Netherland}

\author[0000-0003-4793-7880]{Fabian Walter}
\affil{Max Planck Institute f\"ur Astronomie, K\"onigstuhl 17, 69117 Heidelberg, Germany}
\affil{National Radio Astronomy Observatory, Pete V. Domenici Array Science Center, P.O. Box O, Socorro, NM 87801, USA}

\author{Chris Carilli}
\affil{National Radio Astronomy Observatory, Pete V. Domenici Array Science Center, P.O. Box O, Socorro, NM 87801, USA}

\author{Gerg\"{o} Popping}
\affil{Max Planck Institute f\"ur Astronomie, K\"onigstuhl 17, 69117 Heidelberg, Germany}

\author{Axel Weiss}
\affil{Max-Planck-Institut f\"ur Radioastronomie, Auf dem H\"ugel 69, 53121 Bonn, Germany}

\author{Roberto J. Assef}
\affil{N\'{u}cleo de Astronom\'{\i}a, Facultad de Ingenier\'{\i}a, Universidad Diego Portales, Av. Ej\'{e}rcito 441, Santiago, Chile}

\author{Roland Bacon}
\affil{Univ. Lyon 1, ENS de Lyon, CNRS, Centre de Recherche Astrophysique de Lyon (CRAL) UMR5574, 69230 Saint-Genis-Laval, France}

\author{Franz Erik Bauer}
\affil{Instituto de Astrof\'{\i}sica, Facultad de F\'{\i}sica, Pontificia Universidad Cat\'olica de Chile Av. Vicu\~na Mackenna 4860, 782-0436 Macul, Santiago, Chile}
\affil{Millennium Institute of Astrophysics (MAS), Nuncio Monse{\~{n}}or S{\'{o}}tero Sanz 100, Providencia, Santiago, Chile}
\affil{Space Science Institute, 4750 Walnut Street, Suite 205, Boulder, CO 80301, USA}

\author{Frank Bertoldi}
\affil{Argelander-Institut f\"ur Astronomie, Universit\"at Bonn, Auf dem H\"ugel 71, 53121 Bonn, Germany}

\author{Richard Bouwens}
\affil{Leiden Observatory, Leiden University, PO Box 9513, NL2300 RA Leiden, The Netherland}

\author{Thierry Contini}
\affil{Institut de Recherche en Astrophysique et Plan\'etologie (IRAP), Universit\'e de Toulouse, CNRS, UPS, 31400 Toulouse, France}

\author{Paulo C.~Cortes}
\affil{Joint ALMA Observatory - ESO, Av. Alonso de C\'ordova, 3104, Santiago, Chile}
\affil{National Radio Astronomy Observatory, 520 Edgemont Rd, Charlottesville, VA, 22903, USA} 

\author{Pierre Cox}
\affil{Institut d'Astrophysique de Paris, 98 bis boulevard Arago, 75014 Paris, France}

\author{Elisabete da Cunha}
\affil{Research School of Astronomy and Astrophysics, Australian National University, Canberra, ACT 2611, Australia}

\author{Emanuele Daddi}
\affil{Laboratoire AIM, CEA/DSM-CNRS-Universite Paris Diderot, Irfu/Service d'Astrophysique, CEA Saclay, Orme des Merisiers, 91191 Gif-sur-Yvette cedex, France}

\author{Tanio D\'iaz-Santos}
\affil{N\'{u}cleo de Astronom\'{\i}a, Facultad de Ingenier\'{\i}a, Universidad Diego Portales, Av. Ej\'{e}rcito 441, Santiago, Chile}

\author{David Elbaz}
\affil{Laboratoire AIM, CEA/DSM-CNRS-Universite Paris Diderot, Irfu/Service d'Astrophysique, CEA Saclay, Orme des Merisiers, 91191 Gif-sur-Yvette cedex, France}

\author{Jacqueline Hodge}
\affil{Leiden Observatory, Leiden University, PO Box 9513, NL2300 RA Leiden, The Netherland}

\author{Hanae Inami}
\affil{Hiroshima Astrophysical Science Center, Hiroshima University, 1-3-1 Kagamiyama, Higashi-Hiroshima, Hiroshima 739-8526, Japan}

\author{Rob Ivison}
\affil{European Southern Observatory, Karl-Schwarzschild-Strasse 2, 85748, Garching, Germany}
\affil{Institute for Astronomy, University of Edinburgh, Royal Observatory, Blackford Hill, Edinburgh EH9 3HJ}

\author{Olivier Le F\`evre}
\affil{Aix Marseille Universit\'e, CNRS, LAM (Laboratoire d'Astrophysique de Marseille), UMR 7326, F-13388 Marseille, France}

\author{Benjamin Magnelli}
\affil{Argelander-Institut f\"ur Astronomie, Universit\"at Bonn, Auf dem H\"ugel 71, 53121 Bonn, Germany}


\author{Pascal Oesch}
\affil{Department of Astronomy, University of Geneva, Ch. des Maillettes 51, 1290 Versoix, Switzerland}
\affil{International Associate, Cosmic Dawn Center (DAWN) at the Niels Bohr Institute, University of Copenhagen and DTU-Space, Technical University of Denmark}

\author[0000-0001-9585-1462]{Dominik Riechers}
\affil{Cornell University, 220 Space Sciences Building, Ithaca, NY 14853, USA}
\affil{Max Planck Institute f\"ur Astronomie, K\"onigstuhl 17, 69117 Heidelberg, Germany}



\author{Ian Smail}
\affil{Centre for Extragalactic Astronomy, Department of Physics, Durham University, South Road, Durham, DH1 3LE, UK}

\author{Rachel S. Somerville}
\affil{Department of Physics and Astronomy, Rutgers, The State University of New Jersey, 136 Frelinghuysen Rd,
Piscataway, NJ 08854, USA}
\affil{Center for Computational Astrophysics, Flatiron Institute, 162 5th Ave, New York, NY 10010, USA}

\author{A. M.  Swinbank}
\affil{Centre for Extragalactic Astronomy, Department of Physics, Durham University, South Road, Durham, DH1 3LE, UK}

\author{Bade Uzgil}
\affil{National Radio Astronomy Observatory, Pete V. Domenici Array Science Center, P.O. Box O, Socorro, NM 87801, USA}
\affil{Max Planck Institute f\"ur Astronomie, K\"onigstuhl 17, 69117 Heidelberg, Germany}

\author{Paul van der Werf}
\affil{Leiden Observatory, Leiden University, PO Box 9513, NL2300 RA Leiden, The Netherland}

\author{Jeff Wagg}
\affil{SKA Organization, Lower Withington Macclesfield, Cheshire SK11 9DL, UK}


\author{Lutz Wisotzki}
\affil{Leibniz-Institut f\"ur Astrophysik Potsdam (AIP), An der Sternwarte 16, 14482 Potsdam, Germany}

%
\begin{abstract}

We analyze the interstellar medium properties of a sample of sixteen bright CO line emitting galaxies identified in the ALMA Spectroscopic Survey in the {\it Hubble} Ultra Deep Field (ASPECS) Large Program. This CO$-$selected galaxy sample is complemented by two additional CO line emitters in the UDF that are identified based on their MUSE optical spectroscopic redshifts. The ASPECS CO$-$selected galaxies cover a larger range of star-formation rates and stellar masses compared to literature CO emitting galaxies at $z>1$ for which scaling relations have been established previously. Most of ASPECS CO-selected galaxies follow these established relations in terms of gas depletion timescales and gas fractions as a function of redshift, as well as the star-formation rate-stellar mass relation (`galaxy main sequence'). However, we find that $\sim30\%$ of the galaxies (5 out of 16) are offset from the galaxy main sequence at their respective redshift, with $\sim12\%$ (2 out of 16) falling below this relationship. Some CO-rich galaxies exhibit low star-formation rates, and yet show substantial molecular gas reservoirs, yielding long gas depletion timescales.  Capitalizing on the well-defined cosmic volume probed by our observations, we measure the contribution of galaxies above, below, and on the galaxy main sequence to the total cosmic molecular gas density at different lookback times. We conclude that main sequence galaxies are the largest contributor to the molecular gas density at any redshift probed by our observations (z$\sim$1$-$3). The respective contribution by starburst galaxies above the main sequence decreases from z$\sim$2.5 to z$\sim$1, whereas we find tentative evidence for an increased contribution to the cosmic molecular gas density from the passive galaxies below the main sequence.

\end{abstract} 

\section{Introduction} 

One of the major goals of galaxy evolution studies has been to understand how galaxies transform their gas reservoirs into stars as a function of cosmic time, and how they eventually halt their star-formation activity.  

An important development has been the discovery that most of the star-forming galaxies show a tight correlation between their stellar masses and star-formation rates \citep[SFRs; e.g.,][]{brinchmann04, daddi07, elbaz07, elbaz11, noeske07,peng10,rodighiero10,whitaker12, whitaker14,schreiber15}. Galaxies in this sequence, usually called ``main-sequence'' (MS) galaxies, would form stars in a steady state for $\sim1-2$ billion years and dominate the cosmic star-formation activity. Galaxies above this sequence, forming stars at higher rates for a given stellar mass, are called ``starbursts''; and galaxies below this sequence, are called ``passive'' or ``quiescent'' galaxies. The large gas reservoirs necessary to sustain the star-forming activity along the MS would be provided through a continuous supply from the intergalactic medium and minor mergers \citep{keres05, dekel09}.  As a consequence, the fundamental galaxy parameters (SFRs, stellar masses, gas fractions and gas depletion timescales) are found to be closely related at different redshifts. Galaxies above the MS, have boosted their SFRs typically through a major merger event \citep[e.g.][]{kartaltepe12}.

A critical parameter in the interstellar medium (ISM) characterization has been the specific SFR (sSFR), defined as the ratio between the SFR and stellar mass (SFR/$M_{\rm stars}$), which for a linear scaling between these parameters denotes how far a galaxy is from the MS population at a given redshift and stellar mass. As a result of observations of gas and dust in star-forming galaxies at high redshift in the last decades \citep[for a detailed summary, see ][]{tacconi18, freundlich19}, current studies indicate that there is an increase of the gas depletion timescales and a decrease in the molecular gas fractions with decreasing redshift ($z\sim3$ to 1), and that the gas depletion timescales decrease with increasing sSFR \citep{bigiel08, daddi10a, daddi10b, genzel10, genzel15, leroy13, saintonge11b, saintonge13, saintonge16, santini14, sargent14, papovich16, schinnerer16, scoville16, scoville17, tacconi10, tacconi13, tacconi18, freundlich19, wiklind19}. Finally, after galaxies would have formed most of their stellar mass on and above the MS, they would slow down or even halt star-formation when they exhausted most of their gas reservoirs \citep[e.g.][]{peng10}, bringing them below the MS line.

\begin{figure*} 
\centering 
\includegraphics[scale=0.8]{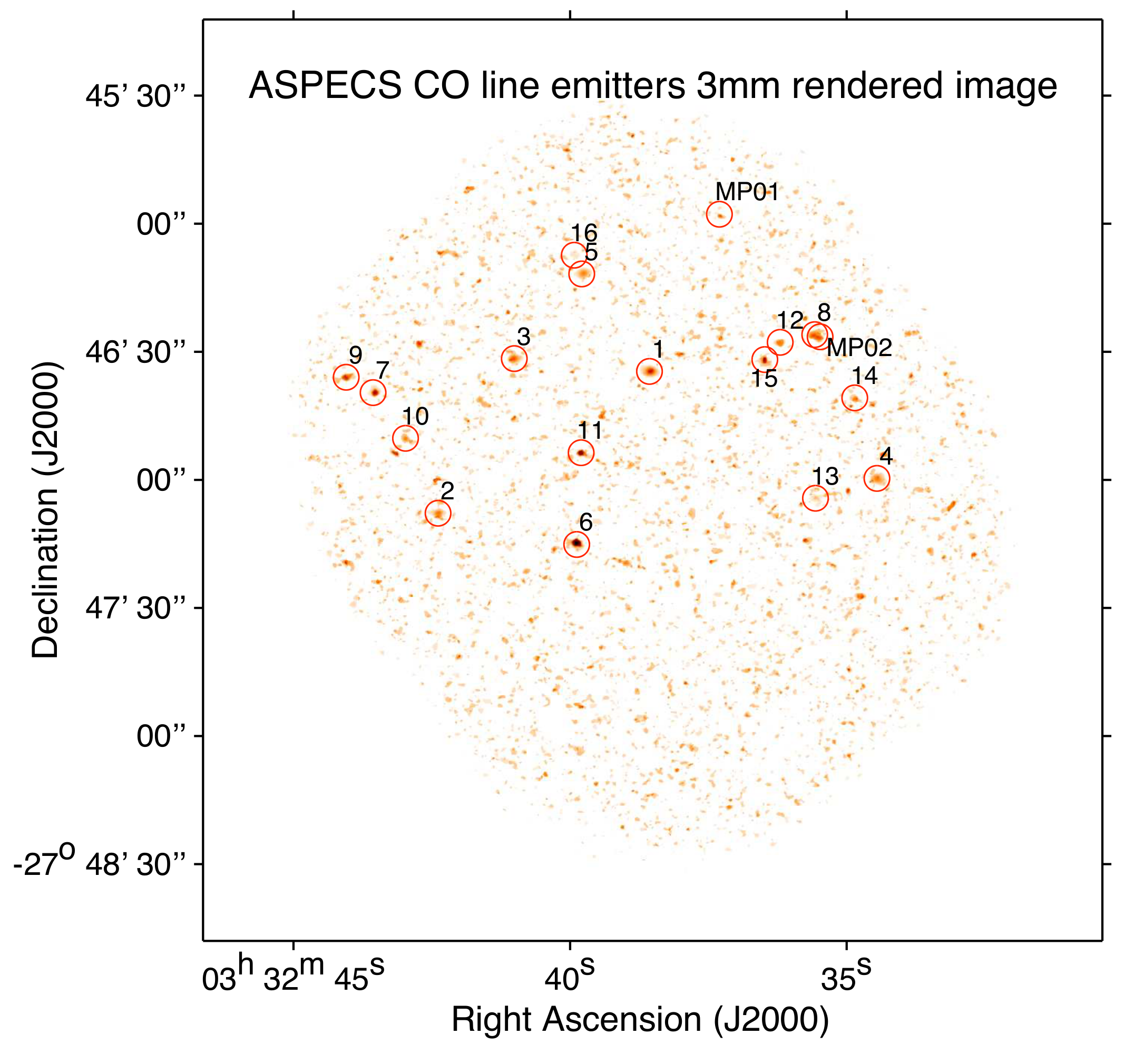}
\caption{Rendered CO image toward the HUDF, obtained by co-adding the individual average CO line maps
around the 16 bright CO-selected galaxies and the 2 lower significance MUSE-based CO sources (labeled MP). Regions with significances below 2.5$\sigma$ in each of the average maps are masked out prior to combination. The location of these individual detections is highlighted by solid circles and their IDs. The tendency of sources to lie in the top two-thirds of the map is likely a combination of clustering and chance, given the sensitivity of the observations is fairly uniform across this region. We note that in this representation of the combined CO map, some individual images might have larger weight (lower noise) than others, and thus some noise peaks might appear as brighter than other statistically significant sources.\label{fig:1}} 
\end{figure*}

Observations of the cold molecular gas in high redshift galaxies have typically relied on transitions of carbon monoxide, $^{12}$CO (hereafter CO), to infer the existence of large gas reservoirs, as CO is the second most abundant molecule in the ISM of star-forming galaxies after H$_2$ and given the difficulty in directly detecting H$_2$ \citep[][]{solomon05, omont07, carilli13}. 

While progress has been substantial, there are still potential biases that have so far been little explored. E. g., most of the high redshift galaxies for which observations of molecular gas and dust are available have been pre-selected from optical and near-IR extragalactic surveys, based on their stellar masses and SFRs estimated from spectral energy distribution (SED) fitting or UV/24$\mu$m photometry. Also due to the finite instrumental bandwidth of millimeter interferometers, CO line studies rely on optical/near-IR redshift measurements. In most cases this means that galaxies need to have relatively bright  emission or absorption lines, or display strong features in the continuum. Similarly, galaxy selection based on detections in {\it Spitzer } and {\it Herschel} far-IR maps, or in ground-based submillimeter observations, will target the most strongly star forming galaxies, and are in many cases affected by source blending due to the poor angular resolution of these space missions. In turn, this means that such source pre-selection will select massive galaxies on or above the massive end of the MS. 

\begin{table*}
\centering
\caption{Observed CO properties\label{table:1}}
\begin{tabular}{cccccccccc}
\hline
ID & RA & Dec & $z_{\rm CO}$ & $J_{\rm up}$ & SNR &  FWHM & $F_{\rm CO}$ & $L'_{\rm CO J\rightarrow(J-1)}$ & $L'_{\rm CO 1-0}$ \\
 & (J2000) & (J2000) &  &  & & (km s$^{-1}$) & (Jy km s$^{-1}$) & ($10^{10}$ K km s$^{-1}$ pc$^2$) & ($10^{10}$ K km s$^{-1}$ pc$^2$)  \\
(1) & (2) & (3) & (4) & (5) & (6) & (7) & (8) & (9 & (10)) \\
\hline
 1 & 03:32:38.54 &$-27$:46:34.62 & 2.543 & 3 & $37.7$ & $517\pm 21$ & $ 1.02\pm 0.04$ & $ 3.40\pm 0.14$ & $ 8.10\pm 0.34$ \\
 2 & 03:32:42.38 & $-27$:47:07.92 & 1.317 & 2 & $17.9$ & $277\pm 26$ & $ 0.47\pm 0.04$ & $ 1.08\pm 0.10$ & $ 1.42\pm 0.13$ \\
 3 & 03:32:41.02 & $-27$:46:31.56 & 2.453 & 3 & $15.8$ & $368\pm 37$ & $ 0.41\pm 0.04$ & $ 1.28\pm 0.12$ & $ 3.04\pm 0.28$ \\
 4 & 03:32:34.44 & $-27$:46:59.82 & 1.414 & 2 & $15.5$ & $498\pm 47$ & $ 0.89\pm 0.07$ & $ 2.31\pm 0.19$ & $ 3.03\pm 0.25$ \\
 5 & 03:32:39.76 & $-27$:46:11.58 & 1.550 & 2 & $15.0$ & $617\pm 58$ & $ 0.65\pm 0.06$ & $ 2.03\pm 0.19$ & $ 2.67\pm 0.24$ \\
 6 & 03:32:39.90 & $-27$:47:15.12 & 1.095 & 2 & $11.9$ & $307\pm 33$ & $ 0.48\pm 0.06$ & $ 0.77\pm 0.09$ & $ 1.01\pm 0.12$ \\
 7 & 03:32:43.53 & $-27$:46:39.47 & 2.697 & 3 & $10.9$ & $609\pm 73$ & $ 0.76\pm 0.09$ & $ 2.81\pm 0.34$ & $ 6.68\pm 0.81$ \\
 8 & 03:32:35.58 & $-27$:46:26.16 & 1.382 & 2 & $ 9.5$ & $ 50\pm  8$ & $ 0.16\pm 0.03$ & $ 0.39\pm 0.06$ & $ 0.52\pm 0.08$ \\
 9 & 03:32:44.03 & $-27$:46:36.05 & 2.698 & 3 & $ 9.3$ & $174\pm 17$ & $ 0.40\pm 0.04$ & $ 1.48\pm 0.16$ & $ 3.52\pm 0.39$ \\
10 & 03:32:42.98 & $-27$:46:50.45 & 1.037 & 2 & $ 8.7$ & $460\pm 49$ & $ 0.59\pm 0.07$ & $ 0.85\pm 0.10$ & $ 1.12\pm 0.13$ \\
11 & 03:32:39.80 & $-27$:46:53.70 & 1.096 & 2 & $ 7.9$ & $ 40\pm 12$ & $ 0.16\pm 0.03$ & $ 0.25\pm 0.05$ & $ 0.33\pm 0.07$ \\
12 & 03:32:36.21 & $-27$:46:27.78 & 2.574 & 3 & $ 7.0$ & $251\pm 40$ & $ 0.14\pm 0.02$ & $ 0.47\pm 0.06$ & $ 1.12\pm 0.15$ \\
13 & 03:32:35.56 & $-27$:47:04.32 & 3.601 & 4 & $ 6.8$ & $360\pm 49$ & $ 0.13\pm 0.02$ & $ 0.42\pm 0.06$ & $ 1.35\pm 0.19$ \\
14 & 03:32:34.84 & $-27$:46:40.74 & 1.098 & 2 & $ 6.7$ & $355\pm 52$ & $ 0.35\pm 0.05$ & $ 0.56\pm 0.08$ & $ 0.73\pm 0.11$ \\
15 & 03:32:36.48 & $-27$:46:31.92 & 1.096 & 2 & $ 6.5$ & $260\pm 39$ & $ 0.21\pm 0.03$ & $ 0.34\pm 0.05$ & $ 0.45\pm 0.07$ \\
16 & 03:32:39.92 & $-27$:46:07.44 & 1.294 & 2 & $ 6.4$ & $125\pm 28$ & $ 0.08\pm 0.01$ & $ 0.18\pm 0.03$ & $ 0.23\pm 0.04$ \\
\hline
MP01 & 03:32:37.30 & $-27$:45:57.80 & 1.096 & 2 & $ 4.5$ & $169\pm 21$ & $ 0.13\pm 0.03$ & $ 0.21\pm 0.05$ & $ 0.28\pm 0.07$ \\
MP02 & 03:32:35.48 & $-27$:46:26.50 & 1.087 & 2 & $ 4.0$ & $107\pm 30$ & $ 0.10\pm 0.03$ & $ 0.16\pm 0.05$ & $ 0.20\pm 0.06$ \\
\hline
\end{tabular}
\\
\flushleft \noindent {\bf Notes.} (1) Source ID. ASPECS-LP.3mm.xx. (2)-(3) CO coordinates of the detection \citep{gonzalezlopez19}. (4) CO redshift. (5) Observed CO transition. (6) Signal to noise ratio of the detection. (7) CO line full width at half maximum (FWHM). (8) Integrated CO line intensity. (9) CO luminosity of the observed CO transition. (10) CO(1-0) luminosity, inferred from the observed transitions, under the assumptions mentioned in the main text.
\end{table*}

A complementary approach to the targeted observations has been the so-called ``molecular line scan'' strategy \citep{carilli02, walter14}. Here, millimeter/centimeter line observations of an extragalactic `blank-field' are performed using a sensitive interferometer, exploring a significant frequency range (e.g. the full 3mm and/or 1mm band) over a sizable area of the sky. This essentially provides a large data cube to search for the redshifted emission from CO emission lines and/or cold dust continuum. Under this approach, galaxies are selected purely based on their molecular gas content. Pioneering observations of the {\it Hubble} Deep Field North (HDF-N) with the Plateau de Bureau Interferometer (PdBI), covering the full 3mm band, led to the first estimates of the CO luminosity functions (LF) at high redshift and the first constraints on the cosmic density of molecular gas \citep{walter14, decarli14}. More recently, observations with the Karl Jansky Very Large Array (VLA) at centimeter wavelengths in the COSMOS field and the HDF-N have allowed to cover larger areas, enabling the characterization of larger samples of gas rich galaxies, and providing tighter constraints on the CO LF and the evolution of the cosmic density of molecular gas \citep{pavesi18, riechers19}. 

The ALMA Spectroscopic Survey (ASPECS) is the first contiguous molecular survey of distant galaxies performed with ALMA. The ASPECS pilot program targeted a region of 1 arcmin$^2$ of the {\it Hubble} Ultra Deep Field (HUDF), scanning the full 3-mm and 1-mm bands. This enabled independent line searches in each band \citep{walter16}, allowing the investigation of a variety of topics including the characterization of CO selected galaxies \citep{decarli16a}, constraints on the CO LF and cosmic density of molecular gas \citep{decarli16b}, derivation of 1-mm continuum number counts and study of the properties of the faintest dusty galaxies \citep{aravena16b, bouwens16}, searches for [CII] line emission at $z>6$ \citep{aravena16c} and derivation of constraints for CO intensity mapping experiments \citep{carilli16}. 

\begin{figure*} 
\centering 
\includegraphics[scale=0.9]{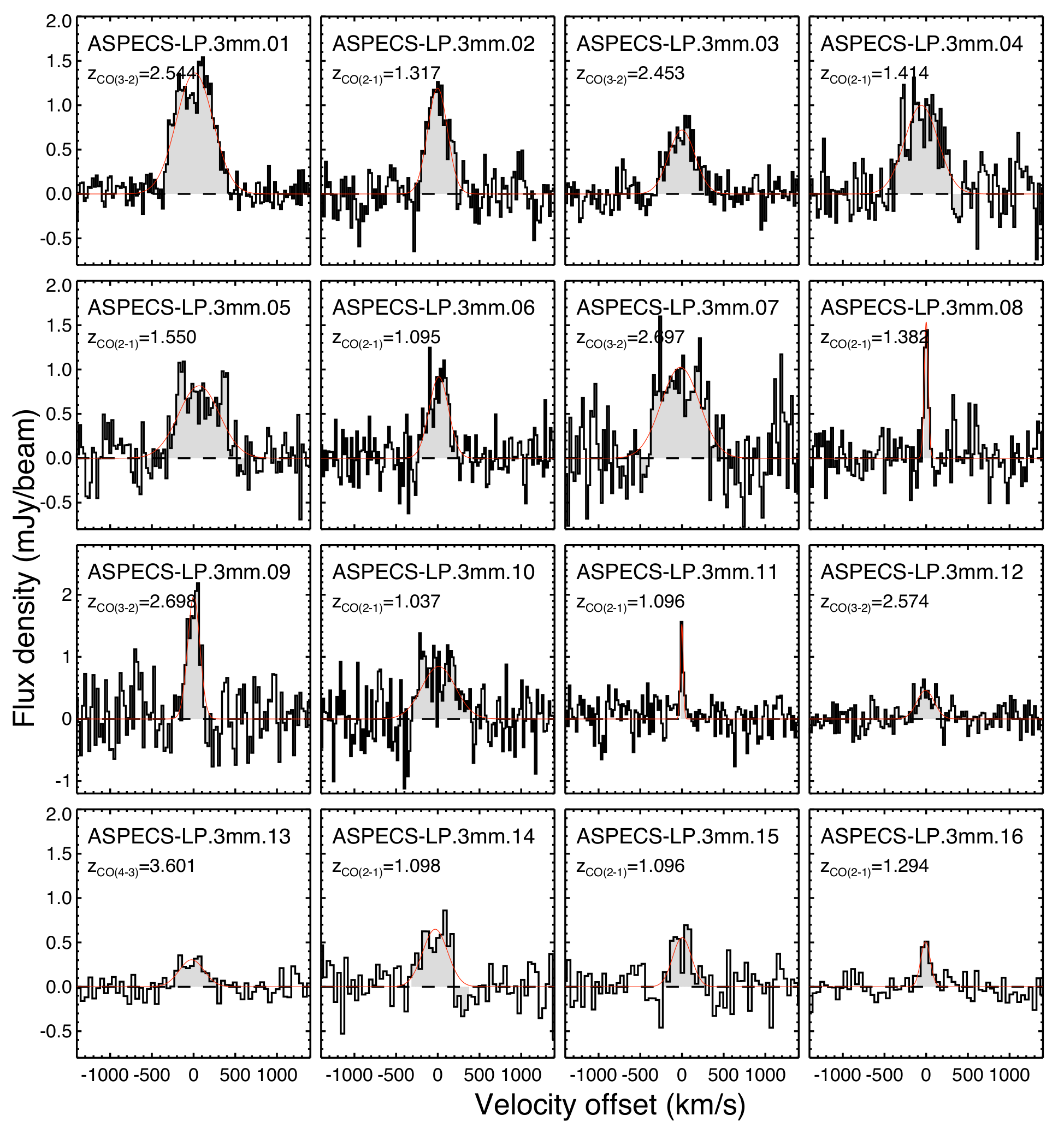}
\caption{CO line emission profiles obtained from the ALMA 3-mm data cube, toward the 16 most significant CO-selected detections. The spectra are centered at the identified line, and shown at a width of 7.813 MHz per channel ($\sim25$ km s$^{-1}$). For the sources in the bottom row, the spectra have been rebinned by a factor of 2. The red solid line, represents a 1-dimensional Gaussian fit to the profiles. The profiles are obtained by extracting the spectra in the original cube, at the location of the peak position identified in the moment-0 image. The grey-shaded area corresponds to the velocity range used to obtain the moment-0 images used in Fig.\ref{fig:1}.\label{fig:2}} 
\end{figure*}

The ASPECS program has since been expanded, representing the first extragalactic ALMA large program (LP). ASPECS LP builds upon the observational strategy and the results presented by the ASPECS pilot observations, but extending the covered area of the HUDF from $\sim1$ arcmin$^2$ to 5 arcmin$^2$, comprising the full area encompassed by the {\it Hubble} eXtremely Deep Field (XDF). We here report results based on the 3mm data obtained as part of the ASPECS LP. 

The ASPECS LP survey strategy and derivation of the CO luminosity function and evolution of the cosmic molecular gas density are presented by \citet{decarli19}. The line and continuum search techniques, as well as 3-mm continuum image and number counts are presented in  \citet{gonzalezlopez19}. The optical source and redshift identification and global galaxy properties, based on the ultra deep optical/near-IR coverage of the UDF are presented in \citet{boogaard19}. A theoretical prediction of the cosmic evolution of the CO luminosity function and comparison to the ASPECS measurements are presented in \citet{popping19}. 

In this paper, we analyze the ISM properties of the 16 statistically reliable CO line identifications plus 2 lower significance CO lines identified through optical redshifts, and compare them with the properties of previous targeted CO observations at high redshift. In Section \ref{sec_obs}, we briefly summarize the ASPECS LP observations and the ancillary data used in this work. In Section \ref{sec_results}, we present the CO line properties. In Section \ref{sec_analysis}, we compare the ISM properties of our ASPECS CO galaxies with standard scaling relations derived from targeted observations of star forming galaxies. In Section \ref{sec_conclusion}, we summarize the main conclusions from this work. Hereafter, we assume a standard $\Lambda$CDM cosmology with $H_0=70$ km s$^{-1}$ Mpc$^{-1}$, $\Omega_\Lambda=0.7$ and $\Omega_M=0.3$. 

\section{Observations} \label{sec_obs}

The ASPECS LP uses the same observational strategy followed by the ASPECS pilot survey \citep{walter16}, but expanding the covered area to $\sim5$ arcmin$^2$. The ASPECS approach is to perform frequency scans over the ALMA bands 3 and 6 (corresponding to the atmospheric bands at 85-115 GHz and 212-272 GHz, respectively) and mapping the selected area through mosaics. The overall strategy is to search in this data cube for molecular gas rich galaxies through their redshifted $^{12}$CO emission lines entering the ALMA bands. 
The ASPECS LP band 3 survey setup and data reduction steps are discussed in detail by \citet{decarli19}. Details about the line search procedures are presented in \citet{gonzalezlopez19}. For completeness, we repeat the most relevant information for the analysis presented here. 

\subsection{ALMA band 3}

ALMA band 3 observations were obtained during Cycle 4 as part of the large program project 2016.1.00324.L. The observations were performed using a 17-point mosaic centered at (R.A., Decl)=(03:32:38.0, $-27$:47:00) in the HUDF. We used the spectral scan mode, covering the ALMA band 3, from 84.0 to 115.0 GHz in 5 frequency setups. This strategy yielded an areal coverage of 4.6 arcmin$^2$ at 99.5 GHz at the half power beam width (HPBW) of the mosaic. Observations were performed in a compact array configuration, C40-3, yielding a synthesized beam of $1.75''\times1.49''$ at 99.5 GHz. 

The data were calibrated and imaged using the {\sc CASA} software, using an independent procedure, which follows the ALMA pipeline closely. The visibilities were inverted using the TCLEAN task. Since no very bright sources are found in the data cube, we used the `dirty' cubes. The data were rebinned to a channel resolution of 7.813 MHz, corresponding to 23.5 km s$^{-1}$ at 99.5 GHz. The final cube reaches a sensitivity of $\sim 0.2$ mJy beam$^{-1}$ per 23.5 km s$^{-1}$ channel, yielding $5\sigma$ CO line sensitivities of $\sim (1.4, 2.1, 2.3)\times10^{9}$ K km s$^{-1}$ pc$^2$ for CO(2-1), CO(3-2) and CO(4-3), respectively \citep{decarli19}.  Our ALMA band 3 scan provides coverage for the redshifted line emission from CO(1-0), CO(2-1), CO(3-2) and CO(4-3) in the redshift ranges $0.003-0.369$, $1.006-1.738$, $2.008-3.107$ and $3.011-4.475$, respectively \citep{walter16, decarli19}.

\subsection{CO sample}

To inspect the data cubes we used the \texttt{LineSeeker} line search routine \citep{gonzalezlopez17}. This algorithm convolves the data along the frequency axis with an expected input line width, reporting pixels with signal to noise (S/N) values above a certain threshold. Kernel widths ranging from 50 to 500 km s$^{-1}$ were adopted.  The probability of each line candidate of not being due to noise peaks, or {\it Fidelity}, $F$, was assessed by using the number of negative line sources in the data cube, with $F=1-N_{\rm Neg}/N_{\rm Pos}$. Here, $N_{\rm Neg}$ and $N_{\rm Pos}$ correspond to the number of negative and positive emission line candidates with a given S/N value in a particular kernel convolution \citep{gonzalezlopez19}. We select the sources for which the {\it fidelity} is above 0.9. This yields 16 selected line candidates. All of them, except two sources have fidelity values of $1.0$.  We find that 3mm.15 and 3mm.16 have fidelity values of 0.99 and 0.92, respectively. All these sources are very unlikely to be false positives, based on the statistics presented by \citet{gonzalezlopez19}. Two other independent line searches were performed using similar algorithms with the \texttt{findclumps} \citep{walter16, decarli16a} and \texttt{MF3D} \citep{pavesi18} codes. All the algorithms coincide in the statistical reliability of these sources. As we mention below, all the selected sources have reliable and matching optical/near-infrared counterparts. The sample of 16 line candidates thus constitutes our primary sample, all of which have S/N$>6.4$. 

Two additional sources were selected based on the availability of an optical spectroscopic redshift and a matching a positive line feature in the ALMA cube at the corresponding frequency. By construction, these sources are selected at lower significance than the CO-selected sources. For more details please refer to \citet{boogaard19}. This makes up a sample of 18 galaxies detected in CO emission by the ASPECS program in band-3.

\begin{figure*} \centering 
\includegraphics[scale=0.6]{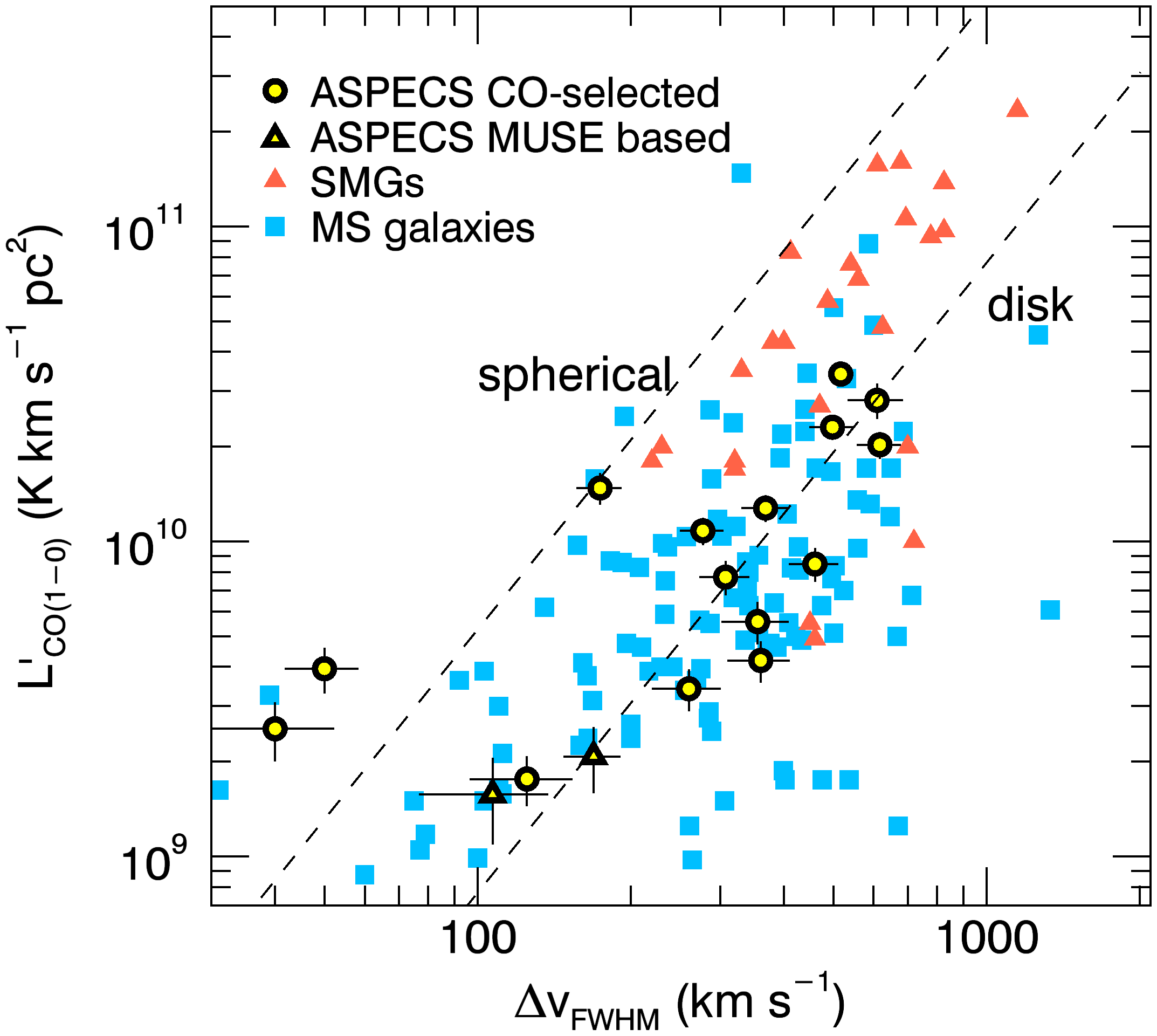}
\includegraphics[scale=0.6]{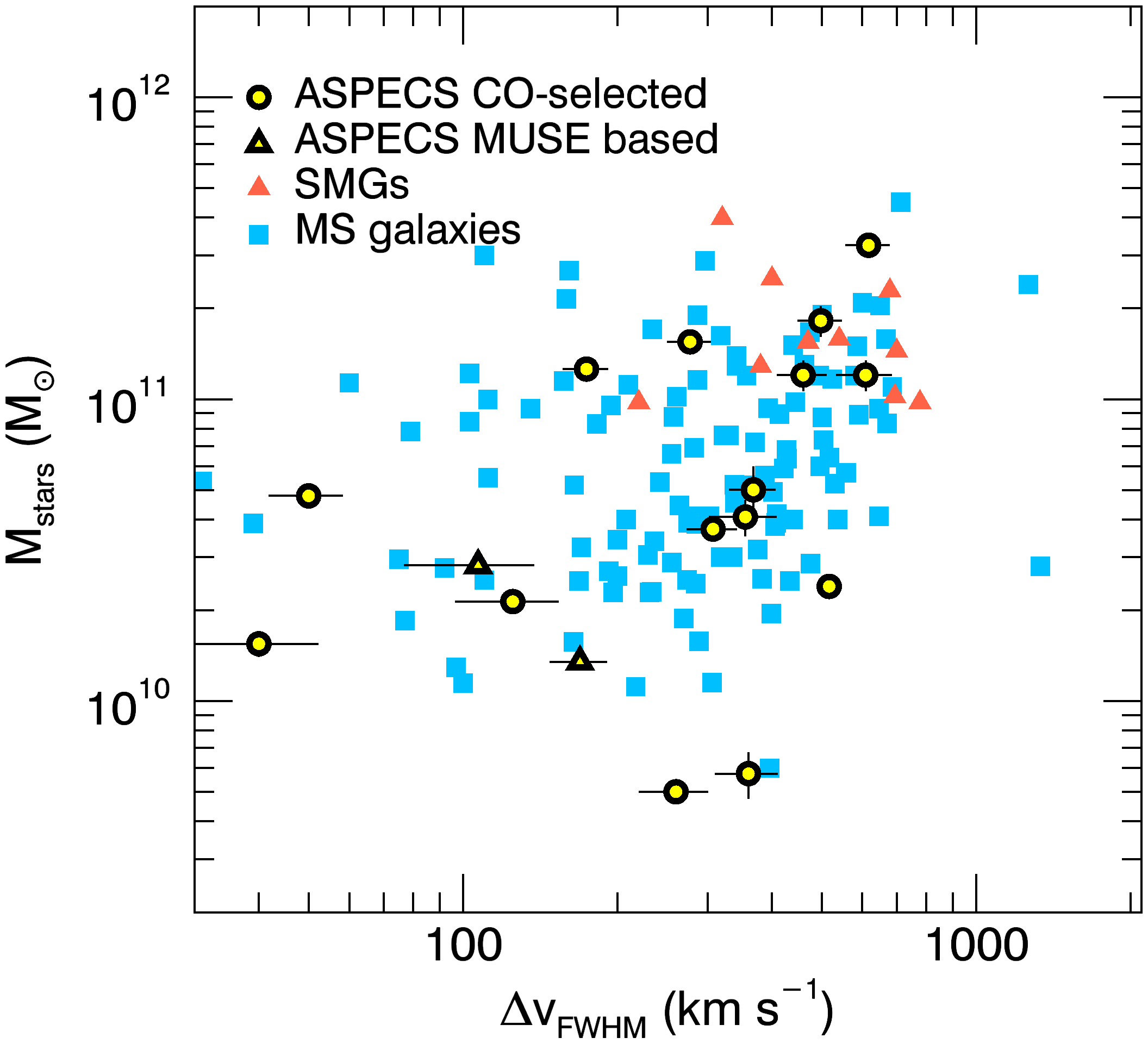}
\caption{({\it Left:}) Estimated CO(1-0) line luminosities as a function of the line widths ($\Delta v_{\rm FWHM}$) for the ASPECS sources, compared to a compilation of galaxies from the literature detected in CO line emission, including unlensed submillimeter galaxies \citep{frayer08,coppin10, riechers11, riechers14, ivison13, ivison11,  thomson12, carilli11, hodge13, bothwell13,  debreuck14} and MS galaxies \citep{daddi10a, magnelli12, magdis12, tacconi13, magdis17, freundlich19}. The dashed lines represent a simple ``virial'' functional form for the CO luminosity for a compact starburst and an extended disk (Sect. \ref{sect:lco_fwhm}). The actual location of each of these lines depend on the choice of geometry and $\alpha_{\rm CO}$ factor. ({\it Right:}) Stellar masses versus line widths for the ASPECS sources, compared to literature (where stellar mass estimates are available). \label{fig:lco_fwhm}} \end{figure*}

\subsection{Ancillary data and SEDs}

Our ALMA observations cover roughly the same region as the {\it Hubble} XDF. Available data include {\it Hubble} Space Telescope ({\it HST}) Advanced Camera for Surveys (ACS) and Wide Field Camera 3 (WFC3) IR data from the HUDF09, HUDF12, and Cosmic Assembly Near-infrared Deep Extragalactic Legacy Survey (CANDELS) programs, as well as public photometric and spectroscopic catalogs \citep{coe06, xu07, rhoads09, mclure13, schenker13, bouwens14, skelton14, momcheva15, morris15, inami17}. In this study, we make use of this optical and infrared coverage of the XDF, including the photometric and spectroscopic redshift information available from \citet{skelton14}. The area covered by the ASPECS LP footprint was observed by the MUSE {\it Hubble} Ultra Deep Survey \citep{bacon17}, representing the main optical spectroscopic sample in this area \citep{inami17}. The Multi-Unit Spectroscopic Explorer (MUSE) at the ESO Very Large Telescope provides integral field spectroscopy in the wavelength range $4750-9350 \AA$ of a $3'\times3'$ region in the HUDF, and a deeper $1'x1'$ region which mostly overlaps with the ASPECS field. The MUSE spectroscopic survey provides spectroscopic redshifts for optically faint galaxies at the $\sim30$ magnitude level, and thus very complimentary to our ASPECS survey. In addition to the {\it HST} coverage, a wealth of optical and infrared coverage from ground-based telescopes is available in this field, including the {\it Spitzer} Infrared Array Camera (IRAC) and Multiband Imaging Photometer (MIPS), as well as by the {\it Herschel} PACS and SPIRE photometry \citep{elbaz11}. From this, we created a master photometric and spectroscopic catalog of the XDF region as detailed in \citet{decarli19}, which includes $>30$ bands for $\sim7000$ galaxies, 475 of which have spectroscopic redshifts. 

We fit the SED of the continuum-detected galaxies using the high-redshift extension of {\sc MAGPHYS} \citep{dacunha08, dacunha15}, as described in detail in \citet{boogaard19}. We use the available broad- and medium-band filters in the optical and infrared regimes, from the U band to {\it Spitzer} IRAC 8 $\mu$m, including also the {\it Spitzer} MIPS 24$\mu$m and Herschel PACS 100$\mu$m and 160$\mu$m. We also include the ALMA 1.2-mm and 3.0-mm data flux densities from \citet{dunlop17} and \citet{gonzalezlopez19}; however we note that the optical/infrared data have a much stronger weight given the tighter constraints in this part of the spectra. We do not include {\it Herschel} SPIRE photometry in the fits since its angular resolution is very poor, being almost the size of our target field for some of the IR bands. For each individual galaxy, we perform SED fits to the photometry fixed at the CO redshift.  {\sc MAGPHYS} employs a physically motivated prescription to balance the energy output at different wavelengths. {\sc MAGPHYS} delivers estimates for the stellar mass, star-formation rate (SFR), dust mass, and IR luminosity.  Estimates on the IR luminosity and/or dust mass come from constraints on the dust-reprocessed UV light, which is well sampled by the UV-to-infrared photometry. The derived parameters are listed in Table \ref{table:2}.

\begin{figure*} \centering \includegraphics[scale=0.8]{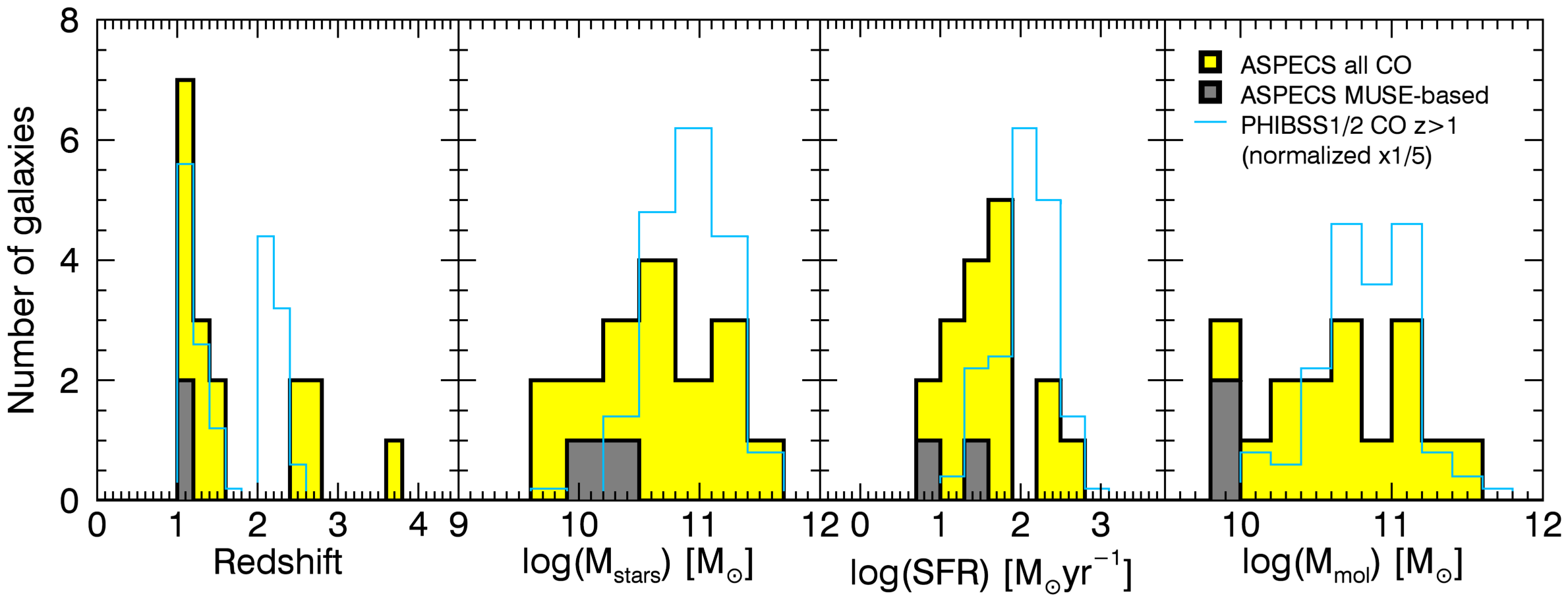}
\caption{Distribution of galaxy properties (SFR, stellar mass, specific SFR and derived gas mass) for the CO line sources in the ASPECS field. The black solid, yellow shaded histograms represent the distributions of all ASPECS CO sources (both CO and MUSE based). The gray shaded histograms present the distribution of the MUSE-based sources only. The light blue histograms show the distribution of the $z>1$ PHIBSS1/2 CO sources \citep{tacconi13,tacconi18}. The number of PHIBSS1/2 sources is normalized by a factor of 1/5 for displaying purposes. Due to its uncertain photometry and thus SED fit, 3mm.12 is not considered in this figure. A fixed conversion factor $\alpha_{\rm CO}=3.6$ (K km s$^{-1}$ pc$^2$)$^{-1}$ has been assumed for the ASPECS CO sources. The comparison sample uses a metallicity-based prescription for this parameter.\label{fig:distr1}} \end{figure*}

\section{Results} \label{sec_results}

\subsection{CO measurements}
 \label{sec_comeasurements}
 
By construction, the ASPECS CO-based sources presented here are selected through their high significance CO line detection. The moment-0 images for each galaxy are created by collapsing the data cube along the frequency axis, considering all the channels within $99.99\%$ percentile range of the line profile. Figure \ref{fig:1} shows a combined CO image of all the moment-0 maps of these sources, highlighting the location of each of these sources in the field. This map is obtained by co-adding all the individual moment-0 line maps, after masking all the pixels with signal to noise ratios below $2.5\sigma$. Figure \ref{fig:2}  shows the CO spectral profiles, obtained at the peak position of each of the sources (see also Appendix \ref{sect:app2}). 

Following \citet{gonzalezlopez19}, the total CO intensities were derived from the ASPECS band-3 data cube by creating moment-0 images, collapsing the cube in velocity around the detected CO lines, and spatially integrating the emission from pixels within a region containing the CO emission \citep[see][for more details]{gonzalezlopez19}. 

All of our CO sources are clearly identified with optical counterparts, as described in detail by \citet{boogaard19}. While for most of these sources a photometric redshift is enough to provide an identification of the actual CO line transition and redshift, a large fraction of them is matched with a MUSE spectroscopic redshift. In three cases (3mm.8, 3mm.12 and 3mm.15), the CO line emission can be either associated to multiple optical sources, due to the higher angular resolution of the optical {\it HST} images, or the candidate CO redshift does not coincide with any of the catalogued photometric or spectroscopic redshifts. In these cases, inspection of the MUSE data cube is critical \citep{boogaard19}. For the source 3mm.13, identified with a CO(4-3) source at $z\sim3.601$ we search for a nearby [CI] 1-0 emission line, however no emission is found at the explored frequency range (see Appendix \ref{sect:app1}). Table \ref{table:1} lists the CO fluxes, positions and derived CO redshifts. 

We compute the CO luminosities, $L'_{\rm CO}$ in units of K km s$^{-1}$ pc$^{2}$, following \citet{solomon97}:

\begin{equation}
L'_{\rm CO}=3.25\times10^7 \nu_{\rm r}^{-2}(1+z)^{-3}D_{\rm L}^2 F_{\rm CO},
\end{equation}
where $\nu_{\rm r}$ is the rest frequency of the observed CO line, in GHz, $D_{\rm L}$ is the luminosity distance at redshift $z$, in Mpc, and $F_{\rm CO}$ is the integrated CO line flux in Jy km s$^{-1}$. Following \citet{decarli16a}, we convert the CO luminosities observed at transition CO($J\rightarrow J-1$) to the ground transition CO($J=1-0$) assuming a line brightness temperature ratio, $r_{\rm J1}=L'_{{\rm CO }J\rightarrow J-1}/L'_{\rm CO 1-0}$. From previous observations of massive MS galaxies \citep{daddi15}, we adopt $r_{21}=0.76\pm0.09$, $r_{31}=0.42\pm0.07$ and  $r_{41}=0.31\pm0.06$. The uncertainties in $L'_{\rm CO}$ account for the uncertainties in the flux measurements and for the uncertainties due to dispersion in the average $r_{J1}$ values measured by \citet{daddi15}. Since the \citet{daddi15} observations do not measure the CO(4-3) lines, but rather CO(3-2) and CO(5-4), we extrapolate between those two lines \citep[i.e. we follow the same approach as ][]{decarli16b}. We note that so far the Daddi et al. CO excitation measurements are the only ones available for similar galaxies at these redshifts. These measurements yield excitation values that are intermediate between low-excitation scenarios such as the external part of the disk in the Milky Way and higher-excitation thermalized scenarios in the $J=3$ to 5 range. This implies that we would not be too far off in either side, if we relax our excitation assumptions. We thus compute the molecular gas masses, in units of $M_\odot$, as

\begin{equation}
M_{\rm H2} = \alpha_{\rm CO} L'_{\rm CO 1-0} = \frac{\alpha_{\rm CO}}{r_{J1}} L'_{{\rm CO }J\rightarrow J-1},
\end{equation}

where $\alpha_{\rm CO}$ is the CO luminosity to gas mass conversion factor in units $M_\odot$ (K km s$^{-1}$ pc$^2$)$^{-1}$. The value of $\alpha_{\rm CO}$ has been found to vary from galaxy to galaxy locally, and to depend on various properties of the host galaxies including metallicity and galactic environment \citep{bolatto13}. There is a clear dependency of decreasing $\alpha_{\rm CO}$ values with increasing metallicity \citep{wilson95, boselli02, leroy11, schruba12, genzel12}, but there is also a trend with morphology, with lower $\alpha_{\rm CO}$ for compact starbursts \citep{downes98} compared to extended disks such as the Milky Way.  Based on previous observations of massive MS galaxies \citep{daddi10a, daddi15, genzel15}, we assume a value $\alpha_{\rm CO}=3.6$ $M_\sun$ (K km s$^{-1}$ pc$^2$)$^{-1}$. 

To check the reliability of our choice of $\alpha_{\rm CO}$, we performed an independent computation of this parameter using the metallicity-dependent approach detailed in \citet{tacconi18}. This method uses assumptions about the stellar mass-metallicity and the $\alpha_{\rm CO}$-metallicity relations. Using this prescription, we find very homogeneous metallicity-dependent $\alpha_{\rm CO}$ values for the ASPECS CO galaxies. Excluding one source, we find a median of 4.4 $M_\odot$ (K km s$^{-1}$ pc$^2$)$^{-1}$ and a standard deviation of 0.5 $M_\odot$ (K km s$^{-1}$ pc$^2$)$^{-1}$. The excluded source, 3mm.13, however, is an outlier with $\alpha_{\rm CO}\sim13$ $M_\odot$ (K km s$^{-1}$ pc$^2$)$^{-1}$. Given the close to solar metallicities measured in our $z\sim1.5$ ASPECS CO sources \citep{boogaard19}, and for consistency with other papers in this series, in the following we assume a fixed $\alpha_{\rm CO}=3.6$ $M_\sun$ (K km s$^{-1}$ pc$^2$)$^{-1}$. This will yield $<0.1$ dex differences in the molecular gas mass estimates throughout this study with respect to the metallicity dependent approach. All the following analysis has been checked to remain unchanged if we were assuming a metallicity-dependent $\alpha_{\rm CO}$ prescription. The computed CO luminosities are listed in Table \ref{table:1}. The corresponding molecular gas masses are listed in Table \ref{table:2}.

\subsection{CO luminosity vs. FWHM}
\label{sect:lco_fwhm}
\begin{figure} 
\centering 
\includegraphics[scale=0.6]{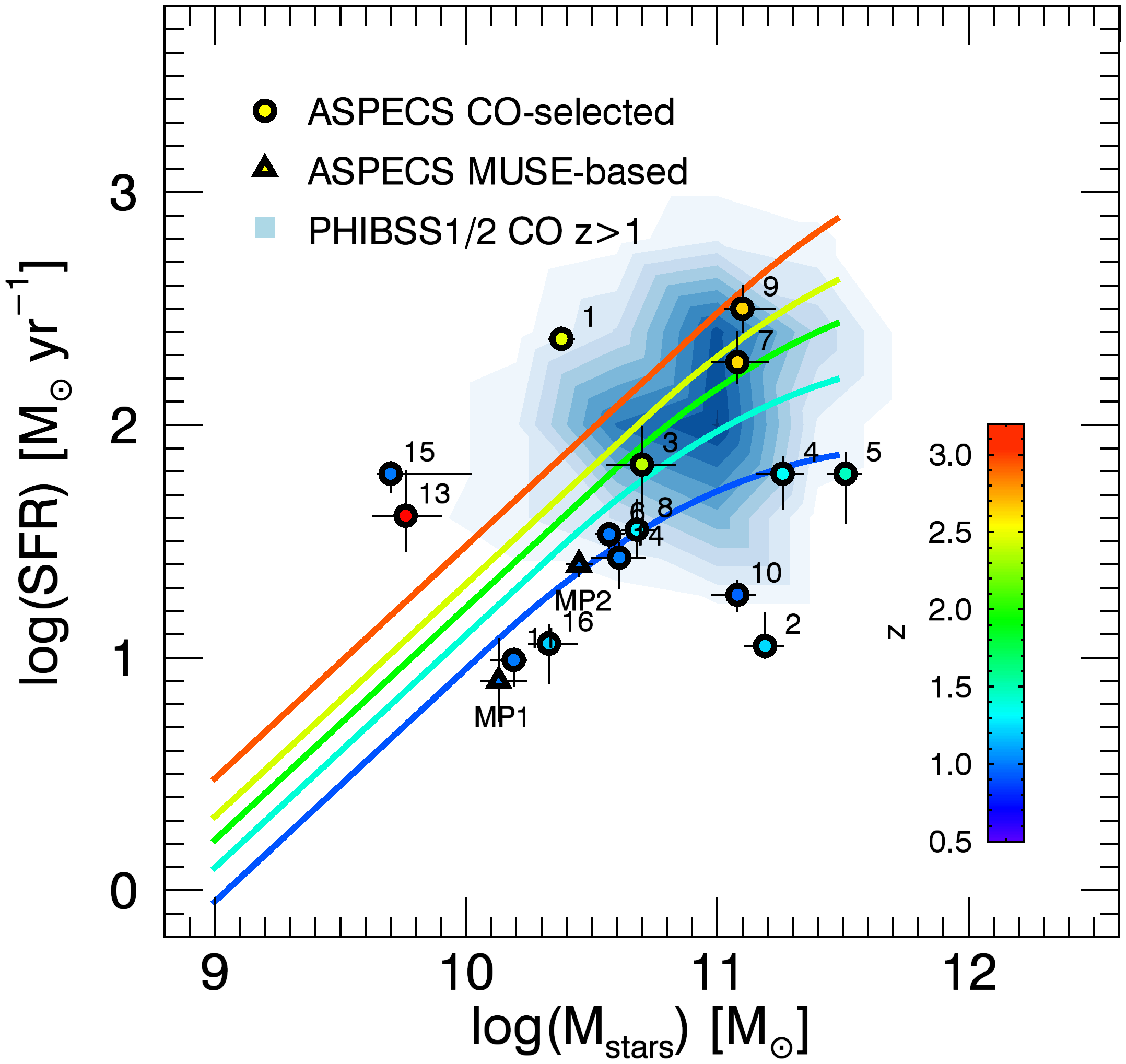}

\caption{SFR vs stellar mass diagram for the ASPECS CO sources, compared to PHIBSS1/2 CO sources at $z>1$. The PHIBSS1/2 galaxies are represented by the blue contours. The solid lines represent the observational relationships between SFR and stellar mass at different redshifts derived by \citet{schreiber15}. These redshifts are denoted in different colors as shown by the color bar to the right. Three of the ASPECS CO selected galaxies lie $>0.4$ dex below the MS at their respective redshift (3mm.2, 3mm.10). \label{fig:ms}}
\end{figure}

Following \citet{bothwell13}, if the CO line emission is able to trace the mass and kinematics of the galaxy then the CO luminosity ($L'_{\rm CO}$), a tracer of the molecular gas mass and thus proportional to the dynamical mass of the system within the CO radius (where baryons are expected to be dominant), should be related to the CO line FWHM. A simple parametrization for this relationship is given by \citep[see][]{bothwell13, harris12, aravena16a}:
\begin{equation}
L'_{\rm CO}=C  \left(\frac{R}{\alpha_{\rm CO} G}\right)  \left(\frac{\Delta v_{\rm FWHM}}{2.35}\right)^2,
\end{equation}
where $R$ is the CO radius in units of kpc, $\Delta v_{\rm FWHM}$ is the line FWHM in km s$^{-1}$, $\alpha_{\rm CO}$ is the CO luminosity to molecular gas mass conversion factor in units of $M_\sun$ (K km s$^{-1}$ pc$^2$)$^{-1}$ and $G$ is the gravitational constant, and $C$ is a constant that depends on the source geometry and inclination \citep{erb06, bothwell13}. A similar argument follows for the possible relation between stellar mass and line FWHM. 

\begin{figure*} 
\centering 
\includegraphics[scale=0.6]{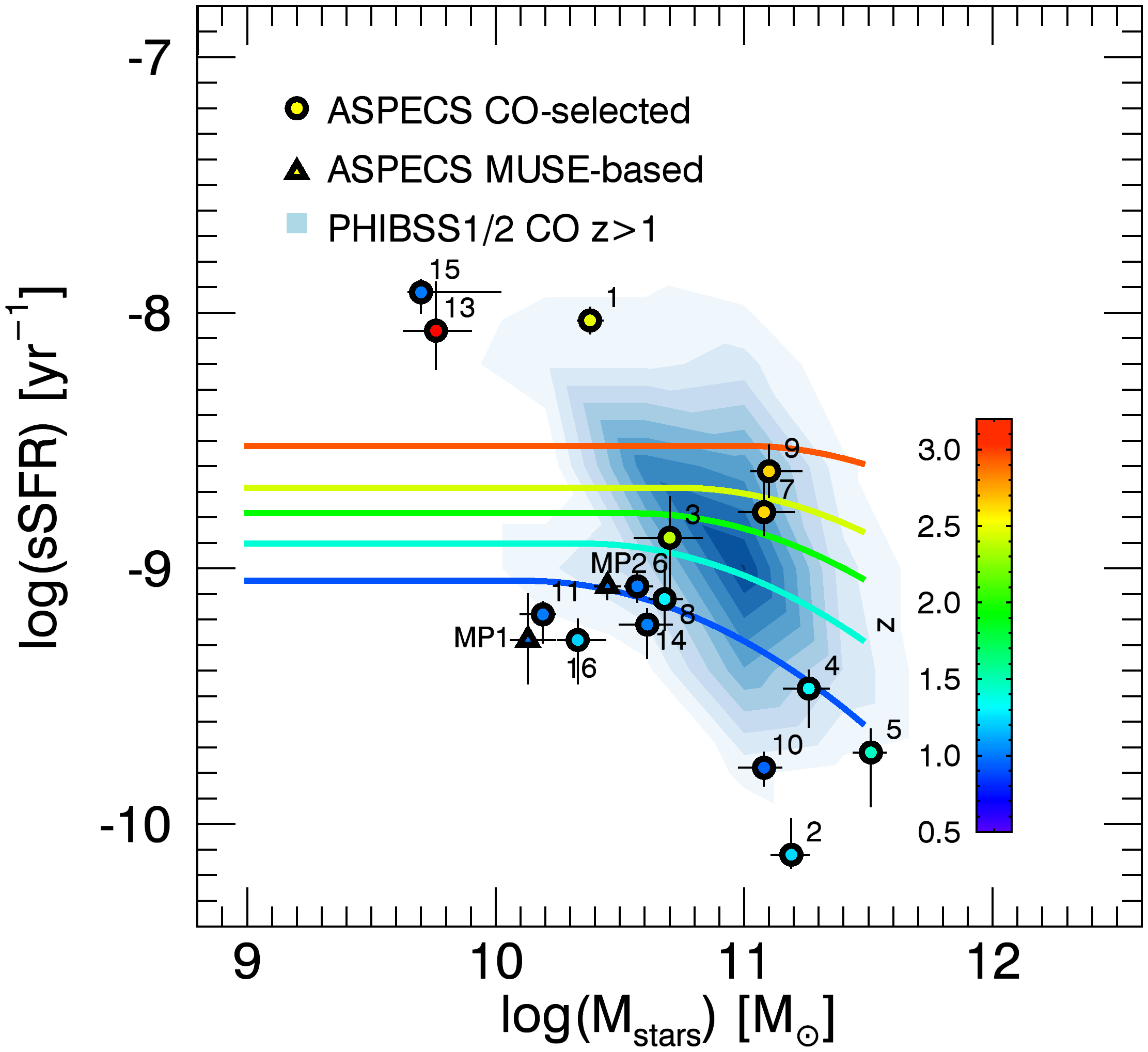}
\includegraphics[scale=0.6]{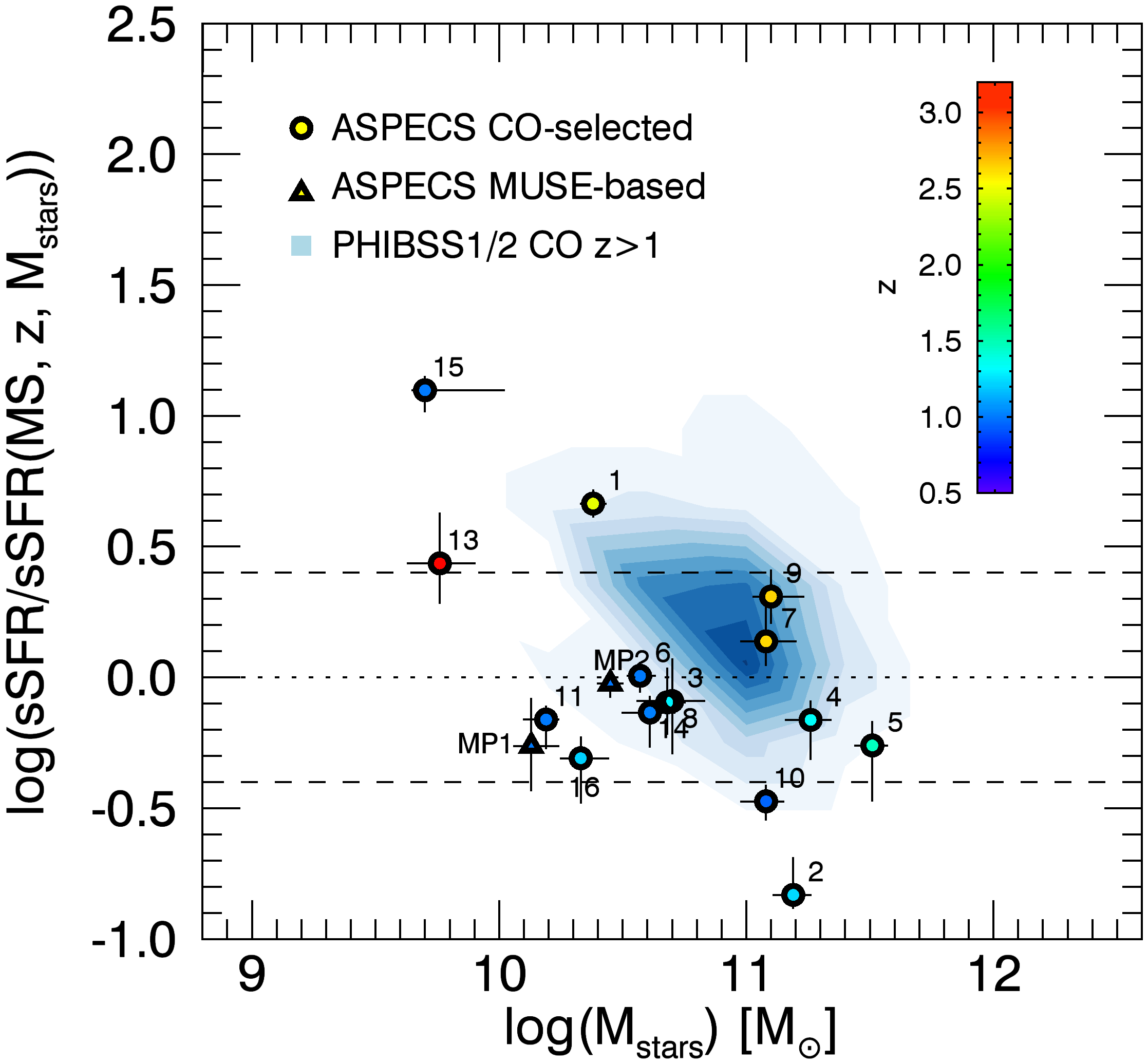}
\caption{({\it Left:}) Specific SFR vs stellar mass diagram for the ASPECS CO sources, compared to $z>1$ PHIBSS1/2 CO sources \citep{tacconi13, tacconi18}. ({\it Right:}) Specific SFR (normalized by the value of the sSFR expected for the MS, which is a function of the redshift and stellar mass) vs stellar mass diagram for the ASPECS CO sources, compared to $z>1$  PHIBSS1/2 CO sources. In both panels, the PHIBSS1/2 galaxies are represented by the blue contours. In the left panel, the solid lines represent the observational relationships between SFR (or sSFR) and stellar mass at different redshifts derived by \citet{schreiber15}. These redshifts are denoted in different colors as shown by the color bar to the right. In the right panel, the dotted line represents the location of the MS, while the dashed lines represents the location of sources at +0.4 and -0.4 dex from the MS. Two of the ASPECS CO selected galaxies lie $>0.4$ dex below the MS at their respective redshift (3mm.2 and 3mm.10). \label{fig:ms2}}
\end{figure*}

Figure \ref{fig:lco_fwhm}-left shows the relationship between the CO luminosities and the line FWHM for our ASPECS CO galaxies, compared to a compilation of high redshift galaxies detected in CO line emission from the literature.  This includes a sample of unlensed submillimeter galaxies \citep{frayer08,coppin10, riechers11, riechers14, ivison13, ivison11,  thomson12, carilli11, hodge13, bothwell13,  debreuck14} and $z>1$ MS galaxies \citep{daddi10a, magnelli12, magdis12, tacconi13, magdis17}. CO line luminosities for MS galaxies have been corrected down to CO(1-0) using the line ratios mentioned above. All the submillimeter galaxies shown have observations of either CO(1-0) or CO(2-1) available, and no correction has been applied in these cases. Also shown in Fig. \ref{fig:lco_fwhm}, are the parametrization of the $L'_{\rm CO}$ vs. FWHM relationship for two representative cases including a disk galaxy model, with $C=2.1$, $R=4$ kpc and $\alpha_{\rm CO}=4.6$ $M_\sun$ (K km s$^{-1}$ pc$^2$)$^{-1}$; and a isotropic (spherical) source, with $C=5$, $R=2$ kpc and $\alpha_{\rm CO}=0.8$ $M_\sun$ (K km s$^{-1}$ pc$^2$)$^{-1}$. A positive correlation is seen between $L'_{\rm CO}$ and the line FWHM, as already found in previous studies \citep[e.g.,][]{bothwell13, harris12, aravena16a}. The scatter in this plot is driven by the different CO sizes ($R$) and inclinations among sources, as well as the choices of $\alpha_{\rm CO}$ and line ratios. Interestingly, most of the ASPECS CO sources seem to cluster around the ``disk'' model line, and would appear that they would follow a preferred geometry. Similarly, most submillimeter galaxies appear to lie closer to the ``spherical'' model line. However, this depends on the choice of parameters for the plotted models (a spherical model would also be able to pass through the ASPECS points).  Inspection of the optical images (see Appendix \ref{sect:app3}) show that the galaxies' morphologies are complex \citep[see also:][]{boogaard19}. Instead, this could either hint toward a possible homogeneity of the ASPECS galaxies in terms of their geometry and $\alpha_{\rm CO}$ factors or just a conspiracy of these. Interestingly, two sources, 3mm.8 and 3mm.11, show very narrow linewidths ($40$ and $50$ km s$^{-1}$, respectively) for their expected $L'_{\rm CO}$. Inspection of the  HST images (see Appendix \ref{sect:app3}) shows that these galaxies are very likely face-on, and thus the reason for such narrow linewidths.

Figure \ref{fig:lco_fwhm}-right shows the stellar mass versus the CO line FWHM. Among the CO sources from the literature, only those with a stellar mass measurement available are shown. More scatter is apparent in this case, arguing for a relative disconnection between the stellar and molecular components. However, the intrinsic uncertainties and differences in the computation of stellar masses makes this difficult to study with the current data.

\begin{figure}
\centering
\includegraphics[scale=0.6]{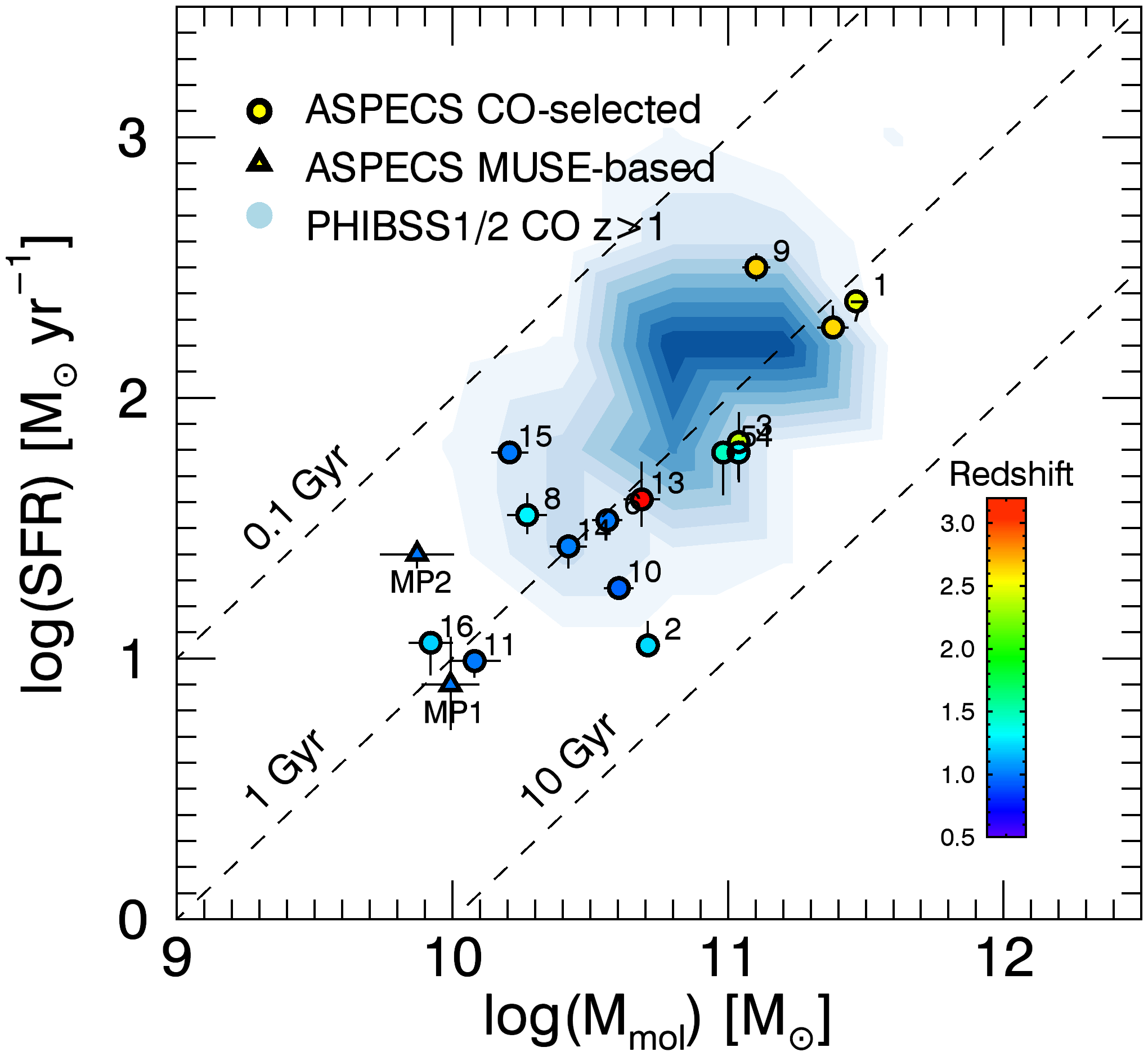}
\caption{SFR vs. $M_{\rm mol}$ for the ASPECS CO galaxies compared to the $z>1$ PHIBSS1/2 CO sources \citep{tacconi13, tacconi18}, represented by blue contours as in Fig. \ref{fig:ms}. The dashed lines represent curves of constant $t_{\rm dep}$ at 0.1, 1 and 10 Gyr. A fixed conversion factor $\alpha_{\rm CO}=3.6$ (K km s$^{-1}$ pc$^2$)$^{-1}$ has been assumed for the ASPECS CO sources. The comparison sample uses a metallicity-based prescription for this parameter. Typical values will range between $\alpha_{\rm CO}=2-5$ (K km s$^{-1}$ pc$^2$)$^{-1}$ for the ASPECS CO sources. \label{fig:ks}}
\end{figure}

\begin{figure*} \centering \includegraphics[scale=0.8]{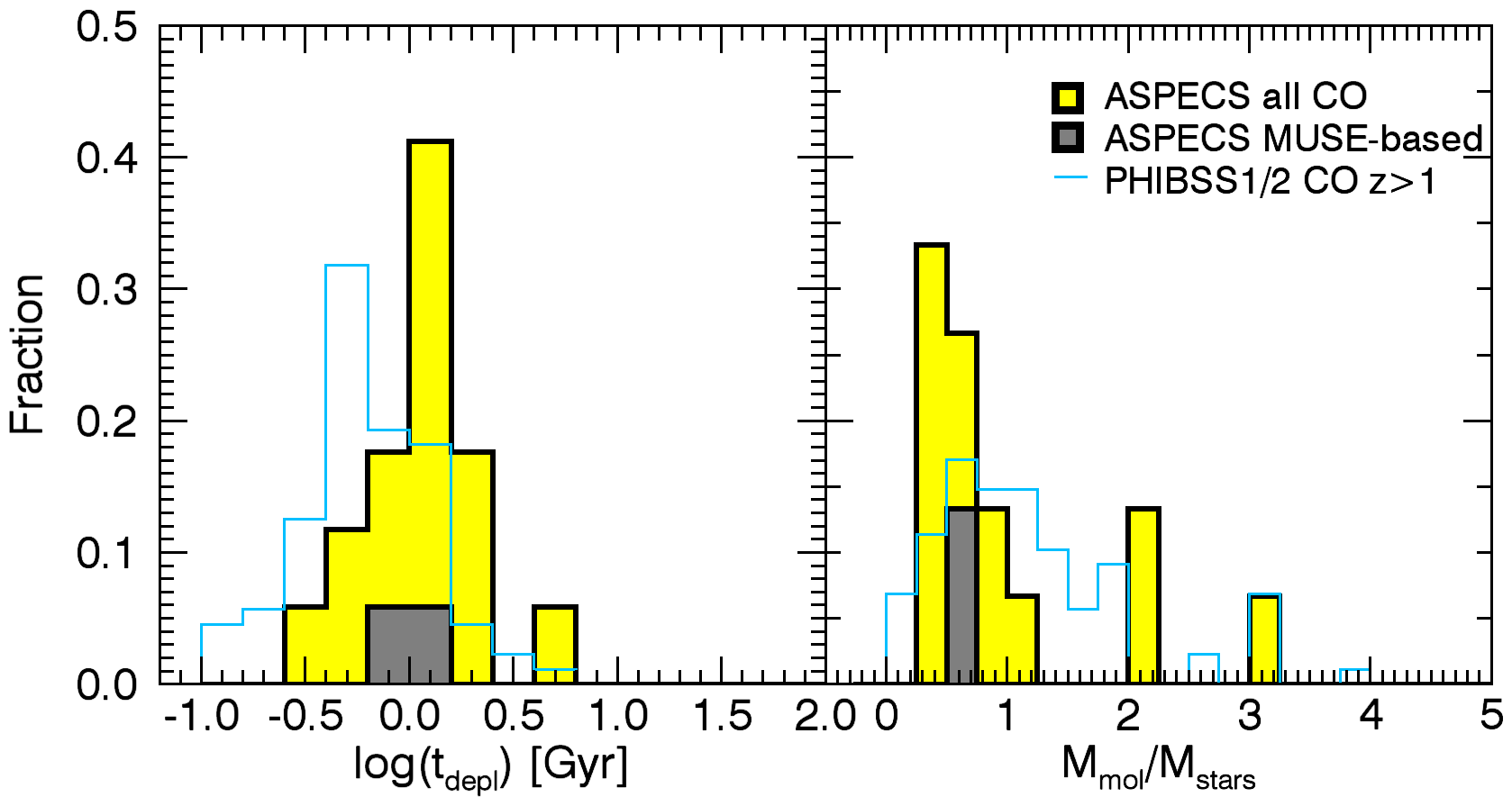}
\caption{Distribution of derived ISM properties (gas depletion timescale and gas fraction) for the CO line sources in the ASPECS field. The black solid, yellow shaded histogram represents the distributions of all ASPECS sources (both CO and MUSE based). The gray shaded histogram show the distribution of the MUSE based sources only. The light blue histograms show the distribution of $z>1$ PHIBSS1/2 CO sources \citep{tacconi13, tacconi18}. Due to its uncertain counterpart photometry, 3mm.12 is not considered in this figure. Sources 3mm.1 and 3mm.13 have high values of $M_{\rm mol}/M_{\rm stars}$ falling outside the range covered by this figure. A fixed conversion factor $\alpha_{\rm CO}=3.6$ (K km s$^{-1}$ pc$^2$)$^{-1}$ has been assumed for the ASPECS CO sources. The comparison sample uses a metallicity-based prescription for this parameter. Typical values will range between $\alpha_{\rm CO}=2-5$ (K km s$^{-1}$ pc$^2$)$^{-1}$ for the ASPECS CO sources.\label{fig:distr2}} 
\end{figure*}

\section{Analysis and Discussion} \label{sec_analysis}

\subsection{CO-selected galaxies in context}

\begin{figure*}
\centering
\includegraphics[scale=0.6]{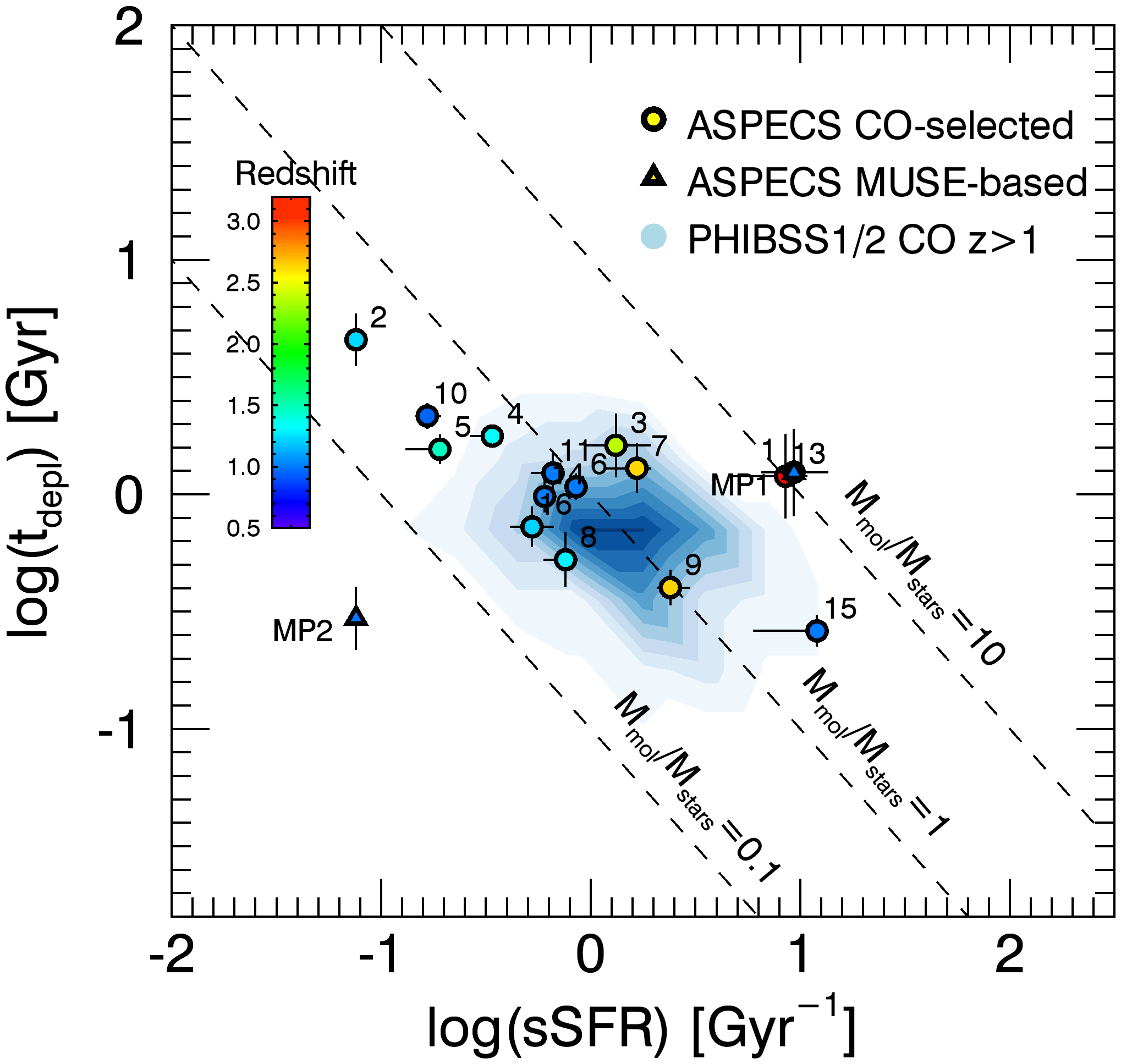}
\includegraphics[scale=0.6]{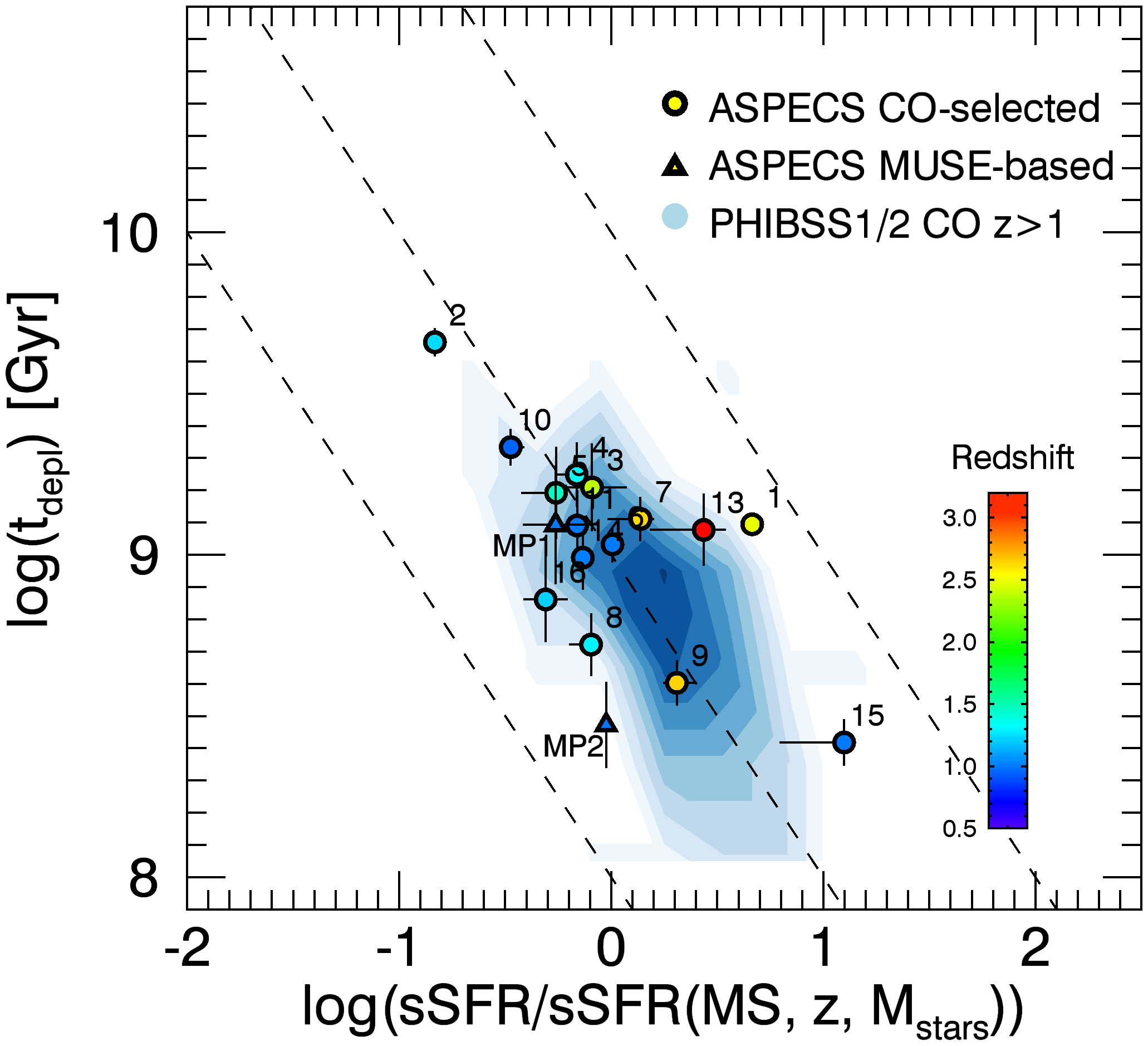}
\caption{The molecular gas depletion timescale ($t_{\rm dep}$) as a function of the specific SFR for the ASPECS CO galaxies. In both panels, the background blue contour levels represent the distribution of $z>1$ PHIBSS1/2 CO galaxies \citep{tacconi13, tacconi18}, and the coloring of each ASPECS source represents their respective redshift. The left panel shows $t_{\rm dep}$ as a function of sSFR. Here the dashed lines represent curves of fixed gas fraction ($M_{\rm mol}/M_{\rm stars}$). The right panel shows the sSFR normalized by the value of the sSFR expected for the MS (which is a function of the redshift and stellar mass) from \citet{schreiber15}. In this case, the dashed lines are shown only for visualization purposes. A fixed conversion factor $\alpha_{\rm CO}=3.6$ (K km s$^{-1}$ pc$^2$)$^{-1}$ has been assumed for the ASPECS CO sources. The comparison sample uses a metallicity-based prescription for this parameter. Typical values will range between $\alpha_{\rm CO}=2-5$ $M_\odot$ (K km s$^{-1}$ pc$^2$)$^{-1}$ for the ASPECS CO sources. \label{fig:tdep_ssfr}}
\end{figure*}

The ASPECS CO survey redshift selection function for CO line detection is roughly limited to galaxies at $z>1$, with a small gap at $z=1.78-2.00$. While it is also possible to detect CO(1-0) for galaxies at $z<0.4$, the volume surveyed is too small to provide enough statistics.

To put our galaxies into context with respect to previous ISM observations, we compare the properties of the ASPECS CO galaxies with the compilation published as part of the ``Plateau de Bureau High-z Blue Sequence Survey'', PHIBBS \citep[][]{tacconi13} and PHIBSS2 \citep{tacconi18, freundlich19}. This provides the largest compilation to date of targeted molecular gas mass measurements from CO line observations, 1-mm dust photometry and far-infrared SEDs for 1444 galaxies selected from different extragalactic fields \citep{saintonge11a, saintonge11b, saintonge16, saintonge17,gao04,graciacarpio08, graciacarpio09,garciaburillo12,bauermeister13,combes11, combes13,tacconi10, tacconi13,genzel15,daddi10a, magdis12,magnelli12,greve05,tacconi06,tacconi08,bothwell13,saintonge13,decarli16b,silverman15,magnelli14,berta16,santini14,bethermin15a,tadaki15, tadaki17,barro16,decarli16b, aravena16b, scoville16, dunlop17,schinnerer16, riechers10b}. The full compilation contains galaxies selected from various different observations and surveys, and thus with different selection functions \citep{tacconi18}. To provide a meaningful comparison, we restrict this sample to sources observed as part of the PHIBSS and PHIBSS2 surveys only, detected in CO line emission at $z>1$ (i.e. exclude dust continuum measurements) from \citet{tacconi18}. This yields a sample of 87 PHIBSS1/2 CO sources at $z>1$, compared to the 18 ASPECS CO sources, spanning a significant range of properties (SFR$\sim10-1000$ M$_\odot$ yr$^{-1}$ and M$_{\rm stars}=10^{9.5}-10^{11.8}$ M$_\odot$).

\begin{figure*} \centering 

\includegraphics[scale=0.65]{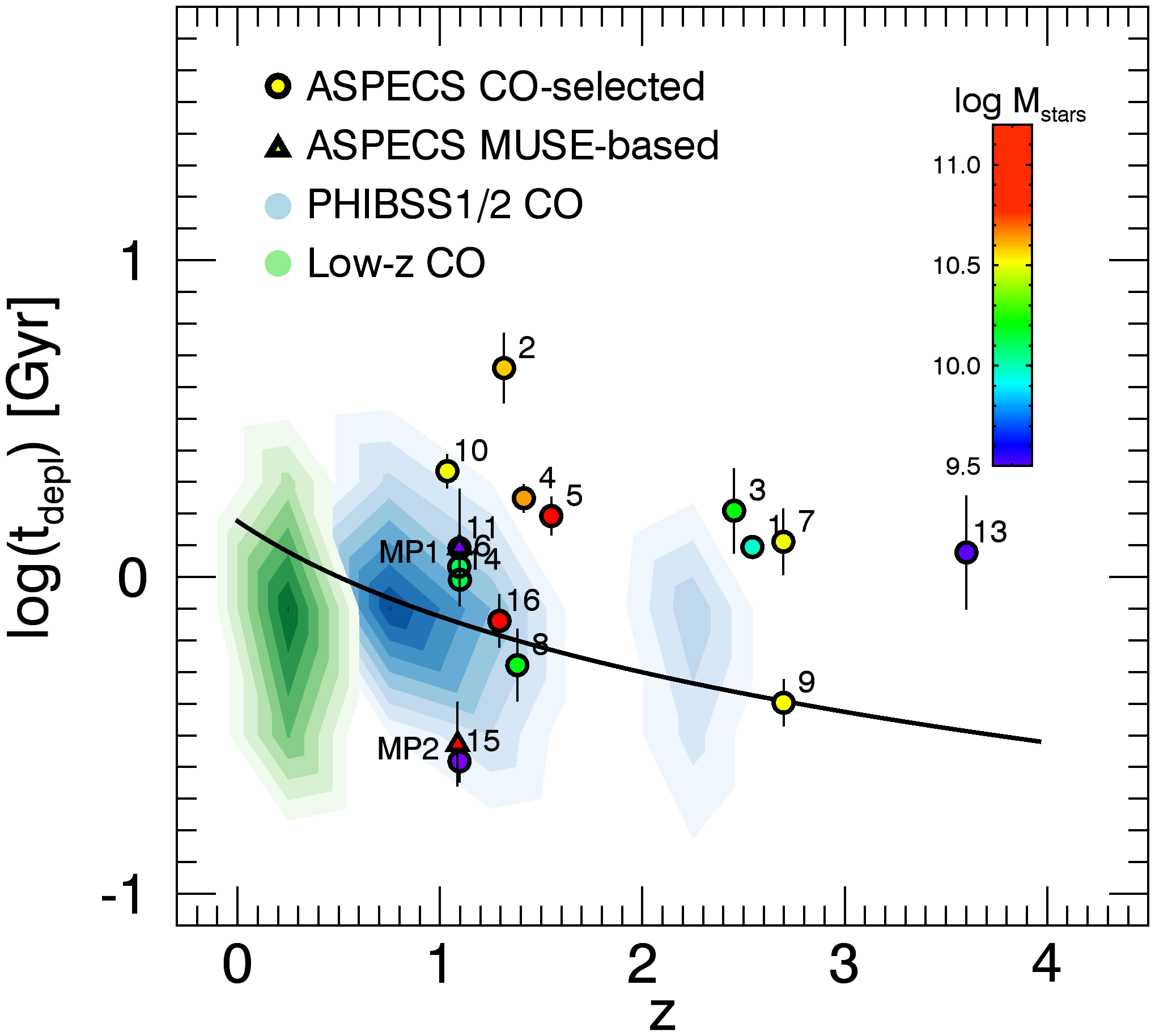}
\includegraphics[scale=0.65]{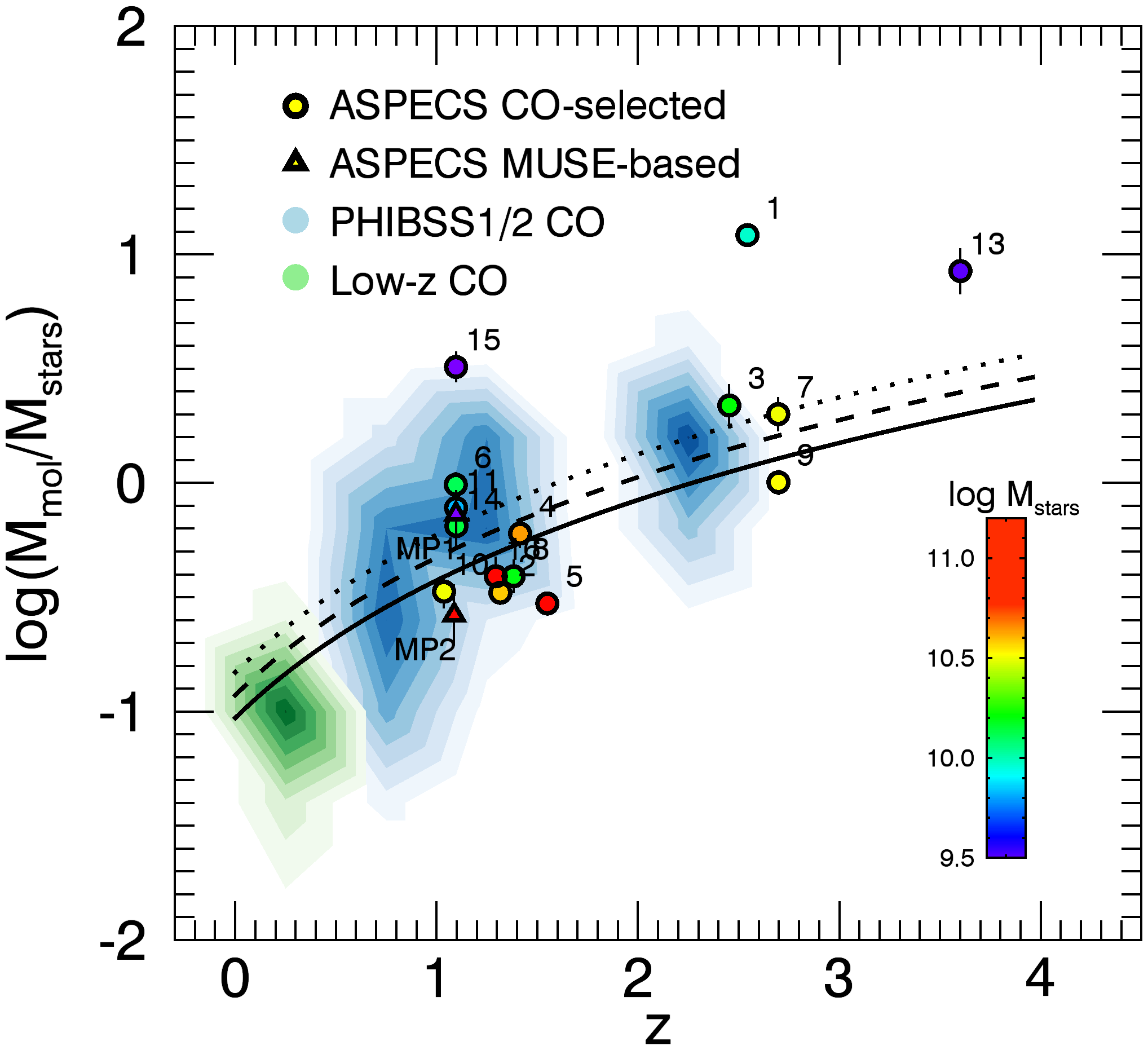} 
\caption{Evolution of the $t_{\rm depl}$ and $f_{\rm mol}=M_{\rm mol}/M_{\rm stars}$ with redshift. The background blue contour levels represent the distribution of galaxies from the PHIBSS1/2 compilation \citep{tacconi13, tacconi18}. As a reference in redshift, we also show as green contours the distribution of galaxies detected in CO line emission at $z<0.5$ from the PHIBSS1/2 compilation (e.g. from xCOLDGASS, GOALS and EgNOG surveys). The solid lines show the expected evolution of  $t_{\rm depl}$ and $f_{\rm mol}$ with redshift, based on previous targeted observations of star forming galaxies. A fixed conversion factor $\alpha_{\rm CO}=3.6$ (K km s$^{-1}$ pc$^2$)$^{-1}$ has been assumed for the ASPECS CO sources. The comparison sample uses a metallicity-based prescription for this parameter. Typical values will range between $\alpha_{\rm CO}=2-5$ $M_\odot$ (K km s$^{-1}$ pc$^2$)$^{-1}$ for the ASPECS CO sources. \label{fig:zevol}} 
\end{figure*}

Given the different nature of the ASPECS survey compared to targeted observations, it is interesting to check how different is the ASPECS selection in terms of basic galaxy parameters. Figure \ref{fig:distr1} shows the distribution of redshift, stellar mass, SFR and CO derived gas masses for all ASPECS CO galaxies, as well as the MUSE based CO sample, compared with the normalized distribution of $z>1$ PHIBSS1/2 CO galaxies (a normalization factor of 1/5 has been used). 

Except for the redshift, these parameters show different distributions for the ASPECS CO galaxies when compared to the $z>1$ PHIBSS1/2 CO galaxies. The ASPECS CO galaxies span a range of two orders of magnitude in stellar mass and three orders of magnitude in SFR. The ASPECS CO galaxies' distributions tend to have lower stellar masses and lower SFRs, with median values of $\sim10^{10.6}\ M_\sun$ and 35 $M_\sun$ yr$^{-1}$, respectively, whereas the bulk of the $z>1$ PHIBSS1/2 CO galaxies have median stellar masses and SFRs of $10^{10.8}$ M$_\odot$ and $\sim$100 M$_\odot$ yr$^{-1}$, respectively. While there are a few literature galaxies with stellar masses below 10$^{10.2}$ M$_\odot$, a larger fraction of ASPECS CO galaxies are located in this range (4 out of 18). We find a clear difference in SFRs between our galaxies and the $z>1$ PHIBSS1/2 CO sample, with all except three ASPECS CO galaxies lying below $\sim100$ $M_\sun$ yr$^{-1}$ and the bulk of the PHIBSS1/2 CO galaxies above this value. Similarly, while almost none of the galaxies in the comparison sample are found with SFR$<25$ M$_\odot$ yr$^{-1}$, five out of the 18 ASPECS CO sources are found in this range. Furthermore, the ASPECS CO galaxies tend to have a flatter distribution of molecular gas masses and some of them show lower values than the PHIBSS1/2 CO galaxies. Since only part of this can be attributed to differences in the assumed $\alpha_{\rm CO}$ factors (as the PHIBSS1/2 survey assumes a metallicity/stellar mass dependent $\alpha_{\rm CO}$), this might reflect differences in parameter space between these surveys, i.e., the lower stellar masses and SFRs inherent to our survey. 

To quantify these differences between the ASPECS CO and the PHIBSS1/2 CO $z>1$ samples, we computed the two sided Kolmogorov Smirnov (KS) statistic, which yields the probability that two datasets are drawn from the same distribution. We find KS probabilities of 0.05, 2.3$\times10^{-4}$ and 0.06 for the stellar mass, SFR and molecular gas mass, respectively. These low values of the KS probability for the stellar mass and SFR distributions point to the differences in the selection between the ASPECS and PHIBSS1/2 surveys, since the latter explicitly did not select galaxies with low SFRs.

Figure \ref{fig:ms} shows the location of the ASPECS CO galaxies in the SFR vs stellar mass plane, compared to the $z>1$ PHIBSS1/2 CO galaxies. The ASPECS galaxies are depicted by large circles and triangles, color-coded to denote their redshifts. Also shown, are the observational relationships derived for the MS galaxies as a function of redshift \citep{schreiber15}. We choose to use the \citet{schreiber15} MS relationships as comparison since this prescription is tunable to a specific redshift, produces curves that are similar to those used in other studies \citep[e.g.,][]{whitaker14, speagle14}, and reproduces the location of the PHIBSS1/2 sources in the MS plane well. A complementary view of the SFR vs stellar mass plot is shown in Fig. \ref{fig:ms2}, which presents the sSFR as a function of the stellar mass. The right panel in particular shows the sSFR normalized by the expected sSFR value of the MS (i.e. the offset from the MS). The sSFR of each galaxy is normalized by the expected sSFR value of the MS at the galaxies' redshift and stellar mass, using the MS prescription presented by \citet{schreiber15}. 

Aside from the larger parameter space explored by the ASPECS survey, as mentioned above, we find two galaxies that are significantly below the MS of star forming galaxies at their respective redshift: 3mm.2 and 3mm.10, corresponding to $12.5\%$ of the CO-selected sample. These galaxies would be classified as `quiescent' galaxies, as their sSFRs are a factor of at least $\sim0.4$ dex below the value of the MS of galaxies at each particular redshift for a fixed stellar mass. Conversely, in three cases (3mm.1, 3mm.13 and 3mm.15) the location of the sources on this plot makes them consistent with `starbursts', {\ bf lying 0.4 dex above the MS, and} corresponding to $18.7\%$ of the CO-selected sample. This implies that $\sim30\%$ of the CO-selected sample corresponds to galaxies off the MS. Note that this would still be valid if we consider systematic uncertainties between different calibrations of the MS as a selection of the MS lines. However, differences in the methods used to compute the SFRs and stellar masses by different studies \citep[e.g.,][]{whitaker14, schreiber15} compared to the {\sc MAGPHYS} SED fitting method used here can bring our `quiescent' sources closer to the respective MS lines \citep[e.g.,][]{mobasher15}. We refer the reader to \citet{boogaard19} for a more detailed discussion on this subject. 

Figure \ref{fig:ks} shows the measured SFRs and CO-derived gas masses for the ASPECS CO galaxies compared to the PHIBSS1/2 CO $z>1$ sample. Dashed lines represent the location of constant depletion timescales ($t_{\rm dep}$; see below for the definition of this parameter). Despite the differences between the ASPECS sources and the PHIBSS1/2 CO $z>1$ sample shown in Figs. \ref{fig:distr1} and \ref{fig:ms}, the majority of the ASPECS galaxies follow relatively tightly the $t_{\rm dep}\sim1$ Gyr line in the SFR$-M_{\rm mol}$ plot (see Fig. \ref{fig:distr2}). This is consistent with the location of the bulk of PHIBSS1/2 CO $z>1$ galaxies, which lie just above this line. Only one ASPECS source, 3mm.2, tend to lie significantly below this trend, closer to the $t_{\rm dep}=10$ Gyr curve. 

Interestingly, we find that the galaxy with the largest offset below the MS line in Fig. \ref{fig:ms}, 3mm.2, appears to have a significant reservoir of molecular gas ($>10^{10}\ M_\sun$), which would be able to sustain star-formation for about 5 Gyr at the current rate (Fig. \ref{fig:ks}). This could be interpreted in the sense that this galaxy might have just recently left the MS of star-forming galaxies and/or might have recently replenished its molecular gas reservoir. Conversely, the starburst galaxies 3mm.9 and 3mm.15 are consistent with short gas depletion timescales ($<1$ Gyr) as typically found in these kind of galaxies.

\begin{table*}
\centering
\caption{ISM properties of ASPECS CO galaxies$^\dagger$.\label{table:2}}
\begin{tabular}{ccccccccc}
\hline

ID & $z_{\rm CO}$ & SFR & $M_{\rm stars}$ & sSFR & $M_{\rm mol}$ & $f_{\rm mol}$ & $t_{\rm dep}$ & $L_{\rm IR}$\\
 &  & ($M_\odot$ yr$^{-1}$) & ($10^{10}\ M_{\odot}$) & (Gyr$^{-1}$) & ($10^{10}\ M_{\odot}$)   &   & (Gyr) & (10$^{11}$ $L_\odot$)\\
(1) & (2) & (3) & (4) & (5) & (6) & (7) & (8) & (9) \\
\hline
 1 & 2.543 & $ 233_{-0}^{+0}$ & $  2.4_{- 0.0}^{+ 0.0}$ & $ 9.3_{-0.0}^{+0.0}$ & $ 29.1\pm  1.2$ & $12.2_{-0.5}^{+0.5}$ & $ 1.2_{-0.1}^{+ 0.1}$ & $80_{- 0}^{+ 0}$ \\
 2 & 1.317 & $  11_{-0}^{+3}$ & $ 15.5_{- 1.0}^{+ 0.7}$ & $ 0.1_{-0.0}^{+0.0}$ & $  5.1\pm  0.5$ & $ 0.33_{-0.04}^{+0.03}$ & $ 4.6_{-0.4}^{+ 1.1}$ & $ 3.1_{- 0.0}^{+ 0.5}$ \\
 3 & 2.453 & $  68_{-20}^{+19}$ & $  5.0_{- 0.9}^{+ 1.0}$ & $ 1.3_{-0.4}^{+0.6}$ & $ 10.9\pm  1.0$ & $ 2.2_{-0.5}^{+0.5}$ & $ 1.6_{-0.5}^{+ 0.5}$ & $ 8.9_{- 2.6}^{+ 2.6}$ \\
 4 & 1.414 & $  61_{-12}^{+ 3}$ & $ 18.2_{- 2.0}^{+ 1.3}$ & $ 0.3_{-0.1}^{+0.0}$ & $ 10.9\pm  0.9$ & $ 0.60_{-0.08}^{+0.07}$ & $ 1.8_{-0.4}^{+ 0.2}$ & $ 9.6_{- 1.2}^{+ 0.2}$ \\
 5 & 1.550 & $  62_{-19}^{+ 6}$ & $ 32_{- 2}^{+1}$ & $ 0.2_{-0.1}^{+0.0}$ & $  9.6\pm  0.9$ & $ 0.30_{-0.03}^{+0.03}$ & $ 1.6_{-0.5}^{+ 0.2}$ & $11_{-3}^{+1}$ \\
 6 & 1.095 & $  34_{-1}^{+  0}$ & $  3.7_{- 0.0}^{+ 0.1}$ & $ 0.9_{-0.0}^{+0.0}$ & $  3.7\pm  0.4$ & $ 1.0_{-0.1}^{+0.1}$ & $ 1.1_{-0.1}^{+ 0.1}$ & $ 3.5_{- 0.1}^{+ 0.0}$ \\
 7 & 2.697 & $ 187_{-16}^{+38}$ & $ 12_{- 1}^{+ 2}$ & $ 1.7_{-0.5}^{+0.3}$ & $ 24\pm  3$ & $ 2.0_{-0.3}^{+0.4}$ & $ 1.3_{-0.2}^{+ 0.3}$ & $22_{-2}^{+4}$ \\
 8 & 1.382 & $  35_{-5}^{+7}$ & $  4.8_{- 0.1}^{+ 0.2}$ & $ 0.8_{-0.2}^{+0.1}$ & $  1.9\pm  0.3$ & $ 0.39_{-0.06}^{+0.06}$ & $ 0.53_{-0.11}^{+ 0.14}$ & $ 4.2_{- 0.6}^{+ 0.8}$ \\
 9 & 2.698 & $ 318_{-34}^{+ 39}$ & $ 13_{-1}^{+3}$ & $ 2.4_{-0.3}^{+0.6}$ & $ 12.7\pm  1.4$ & $ 1.0_{-0.1}^{+0.2}$ & $ 0.40_{-0.06}^{+ 0.07}$ & $36_{-4}^{+4}$ \\
10 & 1.037 & $  18_{-1}^{+0}$ & $ 12._{- 1}^{+1}$ & $ 0.2_{-0.0}^{+0.0}$ & $  4.0\pm  0.5$ & $ 0.33_{-0.05}^{+0.04}$ & $ 2.2_{-0.3}^{+ 0.3}$ & $ 4.5_{- 0.4}^{+ 0.1}$ \\
11 & 1.096 & $  10_{-1}^{+0}$ & $  1.5_{- 0.1}^{+ 0.0}$ & $ 0.7_{-0.1}^{+0.0}$ & $  1.2\pm  0.3$ & $ 0.78_{-0.18}^{+0.16}$ & $ 1.2_{-0.3}^{+ 0.3}$ & $ 1.1_{- 0.1}^{+ 0.0}$ \\
12 & 2.574 & $  31_{-3}^{+18}$ & $  4.4_{- 0.5}^{+ 0.3}$ & $ 0.7_{-0.0}^{+0.5}$ & $  4.1\pm  0.5$ & $ 0.93_{-0.16}^{+0.14}$ & $ 1.3_{-0.2}^{+ 0.8}$ & $ 3.4_{- 0.3}^{+ 2.2}$ \\
13 & 3.601 & $  41_{-8}^{+16}$ & $  0.6_{- 0.1}^{+ 0.1}$ & $ 9_{-4}^{+2}$ & $  4.9\pm  0.7$ & $ 8.5_{-1.9}^{+2.3}$ & $ 1.2_{-0.3}^{+ 0.5}$ & $ 4.2_{- 1.0}^{+ 1.9}$ \\
14 & 1.098 & $  27_{-5}^{+1}$ & $  4.1_{- 0.5}^{+ 0.5}$ & $ 0.6_{-0.00}^{+0.06}$ & $  2.6\pm  0.4$ & $ 0.65_{-0.13}^{+0.12}$ & $ 1.0_{-0.2}^{+ 0.2}$ & $ 3.4_{- 0.8}^{+ 0.2}$ \\
15 & 1.096 & $  62_{-4}^{+0}$ & $  0.5_{- 0.0}^{+ 0.4}$ & $12_{-6}^{+0}$ & $  1.6\pm  0.2$ & $ 3.2_{-0.5}^{+2.8}$ & $ 0.26_{-0.04}^{+ 0.04}$ & $ 6.9_{- 0.0}^{+ 0.0}$ \\
16 & 1.294 & $  11_{-3}^{+1}$ & $  2.1_{- 0.1}^{+ 0.3}$ & $ 0.5_{-0.1}^{+0.1}$ & $  0.8\pm  0.2$ & $ 0.39_{-0.07}^{+0.09}$ & $ 0.73_{-0.22}^{+ 0.14}$ & $ 1.0_{- 0.3}^{+ 0.1}$ \\
\hline
MP01 & 1.096 & $   8_{-2}^{+3}$ & $  1.3_{- 0.1}^{+ 0.2}$ & $ 0.52_{-0.15}^{+0.23}$ & $  1.0\pm  0.2$ & $ 0.73_{-0.17}^{+0.20}$ & $ 1.2_{-0.4}^{+ 0.5}$ & $0_{-80}^{+80}$ \\
MP02 & 1.087 & $  25_{- 0}^{+0}$ & $  2.8_{- 0.0}^{+ 0.0}$ & $ 0.9_{-0.0}^{+0.0}$ & $  0.75\pm0.22$ & $ 0.26_{-0.08}^{+0.08}$ & $ 0.30_{-0.09}^{+ 0.09}$ & $2.9_{-0.2}^{+ 0.7}$ \\
\hline
\end{tabular}
\\
\flushleft \noindent {\bf Notes.} $^\dagger$ As noted by \citet{boogaard19}, formal uncertainties on the derived parameters from the SED fitting are small, systematic uncertanties can be up to 0.3 dex \citep{conroy13}. (1) Source ID. ASPECS-LP.3mm.xx (2) CO redshift. (3)-(5) SFR, stellar mass and specific SFR, derived from {\sc MAGPHYS} SED fitting. (6) Molecular gas mass, computed from the CO line luminosity, $L'_{\rm CO}$ and assuming a CO luminosity to gas mass conversion factor $\alpha_{\rm CO}=3.6$ $M_\odot$ (K km s$^{-1}$ pc$^2$)$^{-1}$. (7) Gas fraction, defined as $f_{\rm mol}=M_{\rm mol}/M_{\rm stars}$. (8) Molecular gas depletion timescale, $t_{\rm dep}=M_{\rm mol}$/SFR. (9) IR luminosity estimate provided by {\sc MAGPHYS} SED fitting.
\end{table*}

\subsection{Gas depletion timescales and gas fractions}

The molecular gas depletion timescale is defined as the time needed to exhaust the current molecular gas reservoir at the current level of star-formation in a galaxy. In the absence of feedback mechanisms (inflows/outflows) the consumption of the molecular gas is driven by star-formation, and thus the gas depletion timescale can be defined as $t_{\rm dep}=M_{\rm mol}/$SFR. Similarly, the gas fraction corresponds to a measurement of how much of the baryonic mass of the galaxy is in the molecular form. This parameter is typically defined as $f_{\rm gas}=M_{\rm mol}/(M_{\rm mol}+M_{\rm stars})$. For this work, we use a simpler quantity, the molecular gas ratio, defined as $\mu_{\rm mol}=M_{\rm mol}/M_{\rm stars}$. Current measurements based on targeted CO and dust observations of star-forming galaxies indicate that both parameters, $t_{\rm dep}$ and $f_{\rm gas}$ (or $\mu_{\rm mol}$), follow clear scaling relations with redshift, sSFR, and stellar mass \citep{scoville17, tacconi18}. These studies indicate that the gas depletion timescales evolve moderately with redshift, following $\propto (1+z)^{\alpha}$.  The value of $\alpha$ has been found to be $-0.62$ from observational studies \citep[e.g.,][]{tacconi13, tacconi18}, while theoretical studies suggest $\alpha=-1.5$ \citep{dave12}. The sSFR follows a steeper evolution with redshift with sSFR$\propto(1+z)^\beta\ M_{\rm stars}^{-0.1}$, with $\beta$ between 5/3 and 3 \citep{lilly13}. Due to the close relationship between these parameters, $\mu_{\rm mol}=t_{\rm dep} {\rm sSFR}$ or $f_{\rm mol}=[1+(t_{\rm dep} {\rm sSFR})^{-1}]^{-1}$, the molecular gas fraction is thus predicted to follow a much stronger evolution with $f_{\rm mol}\propto(1+z)^{1.8 - 2.5}$. To match up the mild evolution of molecular gas depletion timescales with the evolution of the molecular gas ratios or fractions, galaxies might need high accretion rates \citep{scoville17}. While these scaling relations have been successful to describe the properties of star-forming galaxies pre-selected from optical/near-IR surveys, it is not clear to what level they extend to the CO-selected galaxies presented in this study.

Figure \ref{fig:distr2} depicts the distributions of $t_{\rm dep}$ and $\mu_{\rm mol}$ of the ASPECS CO galaxies compared to the $z>1$ PHIBSS1/2 CO galaxies. The range of the distributions of  $t_{\rm dep}$ for both samples appears similar, although the ASPECS CO galaxies seem to have systematically higher $t_{\rm dep}$. This difference could be driven by the lower SFRs in the ASPECS sources and in principle this could be driven by the systematic differences in the SED fitting methods \citep{mobasher15}. However, we should note that some of the ASPECS CO galaxies have systematically lower gas masses. This could be only partly driven by the different prescriptions used for the $\alpha_{\rm CO}$ conversion factor between the different samples, since the distributions of molecular gas masses mostly overlap (Fig. \ref{fig:distr1}). Conversely, the distributions of $\mu_{\rm mol}$ appear similar, covering identical ranges. A KS test comparing the distributions of  $\mu_{\rm mol}$ and $t_{\rm dep}$ yields probabilities of 0.33 and 0.0012, respectively, indicating that the ASPECS CO sources follow a different $t_{\rm dep}$ distribution than the PHIBSS1/2 CO $z>1$ sample. 

Figure \ref{fig:tdep_ssfr} shows the standard scaling relation between $t_{\rm dep}$ and sSFR for the ASPECS CO galaxies, compared to the PHIBSS1/2 CO $z>1$ sample. While the distribution of ASPECS galaxies appears considerably wider in this plane than that of PHIBSS1/2 sources, with a significant fraction of sources having large gas depletion timescales and sSFR below 1 Gyr$^{-1}$, the ASPECS CO galaxies fall well within the lines of constant gas ratio ($M_{\rm mol}/M_{\rm stars}$) at 0.1 and 10 and overall appear to follow the standard relationship between these quantities. This is more clearly seen in the right panel, which shows the sSFR normalized by the expected sSFR value of the MS (i.e. the offset from the MS), using the MS prescription by \citet{schreiber15}. Here, the ASPECS CO-selected galaxies follow the standard linear trend, supporting a direct connection between the distance from the MS and the gas depletion timescale (or inversely the star-formation efficiency). The large span of properties of ASPECS galaxies suggests that a wider parameter space exists beyond that explored by targeted gas/dust observations of pre-selected galaxies.

Figure \ref{fig:zevol} shows the molecular gas depletion timescales and molecular gas ratios of ASPECS CO galaxies as a function of redshift, color-coded by stellar mass, compared to the $z>1$ PHIBSS1/2 CO sample. The ASPECS CO-selected galaxies do not show a particular trend of $t_{\rm dep}$ with redshift, and within the uncertainties they seem consistent with the predicted mild evolution of this parameter. As also shown in Fig. \ref{fig:distr2}, the ASPECS CO galaxies display a significant span in $t_{\rm dep}$ compared to the PHIBSS1/2 sample. A stronger evolution is seen in terms of $M_{\rm mol}/M_{\rm stars}$. If we focus only on the more massive galaxies, depicted as green and red points, there is an obvious increase in the average value of the molecular gas ratio from $M_{\rm mol}/M_{\rm stars}\sim 0.3$ at $z=1$ to $\sim2$ at $z=2.5$. The ASPECS CO-selected sample supports the strong evolution in molecular gas ratio (or fraction) expected from previous targeted observations and models.

\subsection{Molecular gas budget}

Inspection of Fig. \ref{fig:tdep_ssfr} and the color-coding of the data points, suggests there is a tendency of having more starbursting galaxies with increasing redshift (i.e., higher values of sSFR with increasing redshift). Conversely, galaxies tend to be more passive at lower redshifts. This effect is expected by standard scaling relations and has been seen by previous targeted CO surveys \citep[e.g.,][]{tacconi13, tacconi18}. The clean CO-based selection of the ASPECS survey now allows us to investigate how the total budget of molecular gas in galaxies evolves as a function of redshift and distance from the MS (i.e., galaxy type). 

We divided the ASPECS sample into three sets: galaxies significantly above the MS, with log$(\delta_{\rm MS})= log$(sSFR/sSFR$_{\rm MS})$ above 0.4 (``starburst''); galaxies below the MS, with log$(\delta_{\rm MS})<-0.4$ (``passive''); and galaxies within the MS, with $-0.4<$log$(\delta_{\rm MS})<0.4$ (``MS''). We subdivide these samples into two broad redshift bins: $1.0<z<1.7$; $2.0<z<3.1$, which essentially trace the redshift coverage of ASPECS for CO(2-1) and CO(3-2). Each of these redshift bins contains 10 and 4 sources, respectively (sources 3mm.8 was excluded due to the ambiguous optical identification and 3mm.13 due to its redshift outside the defined range). For each redshift bin, we now ask the question of what is the contribution of each galaxy type to the total budget of molecular gas (or what fraction of the total budget they are making up). At each redshift bin, we thus compute this contribution as the sum of all the molecular gas masses from galaxies of this particular type divided by the total molecular gas mass obtained from the recent measurement of cosmic molecular gas density ($\rho_{{\rm H}_2}$) using ASPECS data \citep{decarli19}. 

\begin{figure}
\centering
\includegraphics[scale=0.6]{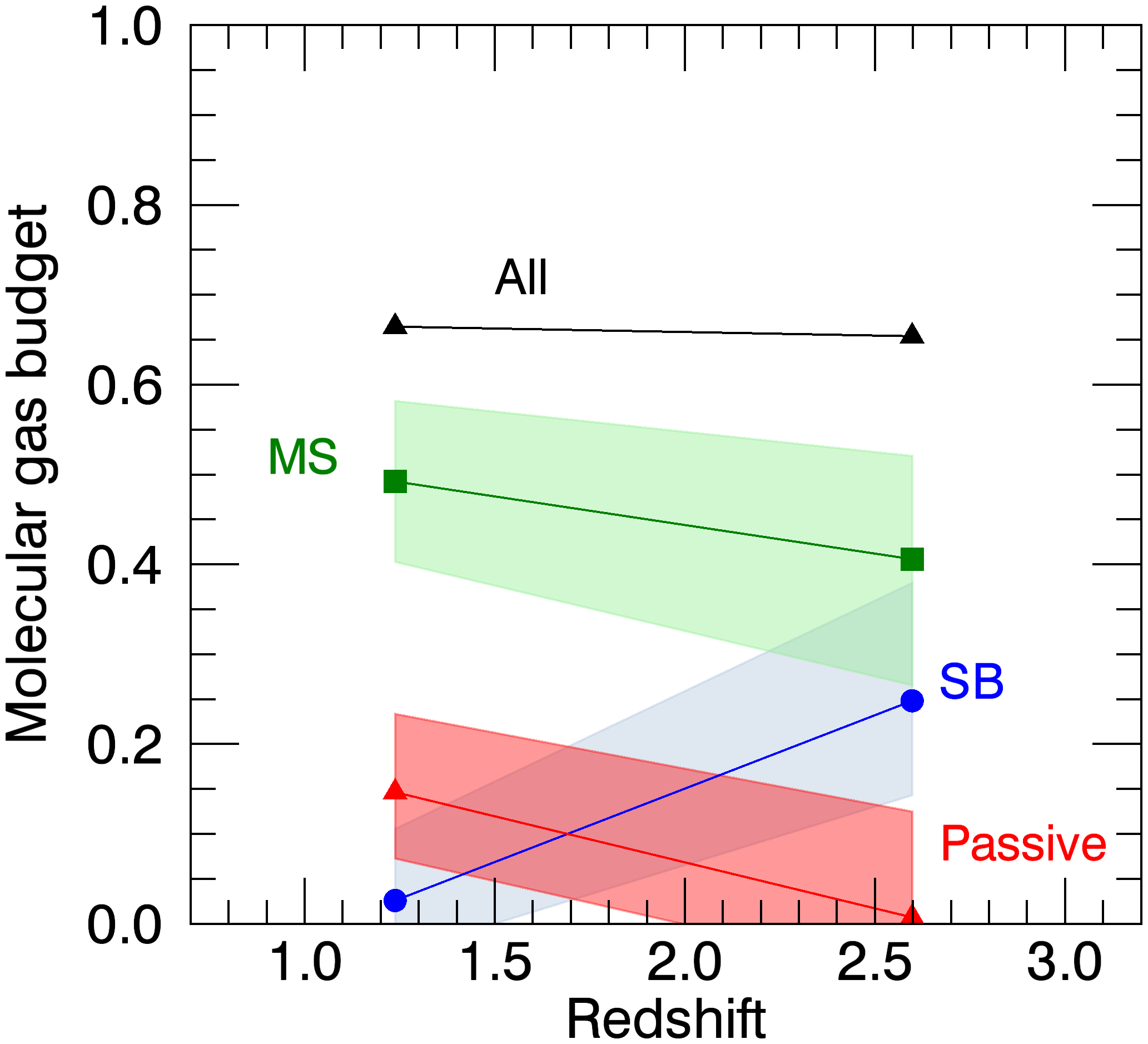}
\caption{Contribution to the total molecular gas budget from galaxies above (starburst), in or below (passive) the MS as a function of redshift inferred from the ASPECS survey. The blue, green and red data points and lines represent galaxies above, in and below the MS, respectively. The black curve shows the contribution of all the CO-selected galaxies considered here to the total molecular gas at each redshift. Each data point is computed from the sum of molecular gas masses of all galaxies in that redshift bin and galaxy type. The redshift measurement of each point is computed as the average redshift from all galaxies in that bin. The shaded region corresponds to the uncertainties of each measurement.\label{fig:budget}}
\end{figure}

The result of this exercise is shown in Fig. \ref{fig:budget}. Here, the different colors represent the galaxy types, and the shaded regions corresponds to the associated uncertainties in these measurements. The values of redshifts used in the horizontal axes correspond to the average redshift among all galaxies in that redshift bin. These uncertainties in the vertical axes are computed as the sum in quadrature of the individual molecular gas mass values, added in quadrature to the statistical uncertainty, which follows binomial distribution, scaled to the total molecular gas in that redshift bin.

The fact that we do not reach full completeness when adding up all CO-selected ASPECS sources is due to the fact that the total molecular gas density also accounts for fainter galaxies that are not part of our sample.

While the analysis is still limited by the admittedly low number of sources (and thus large statistical uncertainties), there appears to be a difference in the trends followed by the different galaxy types. MS galaxies seem to have a dominant contribution to the molecular gas mass budget, which tends to slightly decrease at high redshifts. This decrease, however, is likely driven by the drop in the total contribution from our bright ASPECS galaxies (black curve). Starburst galaxies are consistent with mild evolution, with a contribution increasing from $\sim5\%$ at $z\sim1.2$ to $\sim20\%$ at higher redshift (yet still consistent with no evolution at $1\sigma$). Passive galaxies appear to have a decreasing contribution with increasing redshift, falling from $15\%$ at $z\sim1.2$ to 0$\%$ at $z\sim2.6$. 

Current IR surveys indicate that starburst galaxies have a relatively constant, yet minor, contribution to the cosmic SFR density as a function of redshift, of $\sim8-14\%$ \citep{sargent12,schreiber15}, whereas MS galaxies would have a dominant contribution out to $z=2$. This is consistent with the results presented here in terms of the contribution of starburst and MS galaxies to the molecular gas budget with redshift, and this consistency is expected if the molecular gas content is directly linked to the star formation activity in these kind of galaxies, except only if there is substantial change in efficiencies by a particular galaxy type. However, the decreasing contribution with increasing redshift found for passive galaxies seems to be in contradiction with recent findings by \citet{gobat18} that quiescent early type galaxies at $z=1.8$ have two orders of magnitude more dust than early type galaxies at $z\sim0$. As argued by these authors, this result implies the presence of left-over molecular gas in these $z\sim1.8$ quiescent galaxies, which is consumed in a low-efficient fashion. 

This discrepancy can be understood as follows. Starburst galaxies, typically more abundant at $z>1$, would rapidly exhaust most of their molecular gas reservoirs and typically evolve into passive galaxies. The latter would be more numerous at lower redshifts ($z\sim1$), and might still retain some of the leftover molecular gas from the previous starburst episode(s) \citep[as pointed out by][]{gobat18}. Hence, while passive galaxies might have on average significantly more molecular gas at higher redshifts ($z\sim2-3$), they still represent a very minor fraction of the cosmic molecular gas density or molecular gas budget compared to MS or starburst galaxies. At lower redshifts ($z\sim1$) passive galaxies would have already consumed part of their molecular gas reservoirs, however since they are more numerous, they would contribute an increasing fraction to the cosmic molecular gas density. These ``below MS'' galaxies would thus not only be more prone to be detected by surveys like ASPECS. Perhaps most importantly, this reflects the possibly important, yet overlooked, role of these kind of galaxies in the formation of stars in the universe.

\section{Conclusions} \label{sec_conclusion}

We have presented an analysis of the molecular gas properties of a sample of sixteen CO line selected galaxies in the ALMA Spectroscopic Survey in the {\it Hubble} UDF, plus two additional CO line emitters identified through optical MUSE spectroscopy.

The ASPECS CO-selected galaxies follow a tight relationship in the CO luminosity versus FWHM plane, suggestive of disk like morphologies in most cases. We find that the ASPECS CO galaxies span a range in properties compared to previous pre-selected galaxies with CO/dust follow-up observations. Our galaxies are found to lie at $z\sim1-4$, with stellar masses in the range $0.03-4\times10^{11}\ M_\odot$, SFRs in the range $0-300\ M_\odot$ yr$^{-1}$ and gas masses in the range $5\times10^9 M_\odot$ to $1.1\times10^{11}\ M_\odot$. The wide range of properties shown by the ASPECS CO galaxies expand the range covered by PHIBSS1/2 in CO at $z>1$, with two galaxies falling significantly below the MS ($\sim15\%$) and other three sources ($\sim20\%$) above the MS at their respective redshift. 

The ASPECS CO galaxies are found to tightly follow the SFR-$M_{\rm mol}$ relation, with a typical molecular gas depletion timescale of 1 Gyr, similar to $z>1$ PHIBSS1/2 CO galaxies, yet spanning a range from 0.1 to 10 Gyr. Similarly, the ASPECS sources are found to span a wide range in molecular gas ratios ranging from $M_{\rm mol}/M_{\rm stars}=0.2$ to 6.0. Despite the wide range of properties, the ASPECS CO-selected sources follow remarkably the standard scaling relations trends of $t_{\rm dep}$ and $\mu_{\rm mol}$ with sSFR and redshift.

Finally, we take advantage of the nature of the ASPECS survey to measure the contribution of the molecular gas budget as a function of redshift from galaxies above, in and below the MS. We find a dominant role from MS galaxies. Starburst galaxies appear to have a relatively flat contribution of $\sim10\%$ at $z=1$ and $z=2$. Conversely, passive galaxies appear to have a relevant contribution to the molecular gas budget at $z<1$, yet almost none at $z>1$. We argue this could be due to starburst evolving into passive galaxies at $z\sim1$, and thus an increasing number of passive galaxies with leftover molecular gas.

\acknowledgements We thank Linda Tacconi for making the PHIBSS1/2 compilation available. We also thank the anonymous referee for their helpful and constructive comments. ``Este trabajo cont\'o con el apoyo de CONICYT + Programa de Astronom\'{\i}a+ Fondo CHINA-CONICYT". J.G-L. acknowledges partial support from ALMA-CONICYT project 31160033. F.E.B. acknowledges support from CONICYT grant Basal AFB-170002 (FEB), and the Ministry of Economy, Development, and Tourism's Millennium Science Initiative through grant IC120009, awarded to The Millennium Institute of Astrophysics, MAS (FEB). T.D-S. acknowledges support from ALMA-CONICYT project 31130005 and FONDECYT project 1151239. J.H. acknowledges support of the VIDI research programme with project number 639.042.611, which is (partly) financed by the Netherlands Organizationfor Scientific Research (NWO) D.R. acknowledges support from the National Science Foundation under grant number AST-1614213. I.R.S. acknowledges support from the ERC Advanced Grant DUSTYGAL (321334) and STFC (ST/P000541/1). This paper makes use of the following ALMA data: 2016.1.00324.L. ALMA is a partnership of ESO (representing its member states), NSF (USA) and NINS (Japan), together with NRC (Canada), NSC and ASIAA (Taiwan), and KASI (Republic of Korea), in cooperation with the Republic of Chile. The Joint ALMA Observatory is operated by ESO, AUI/NRAO and NAOJ.

\bibliographystyle{apj} \bibliography{cobright}

\appendix

\section{Search for [CI] line emission}
\label{sect:app1}
The line identification for one of the ASPECS CO detections, 3mm.13, was found consistent with CO(4-3) at a redshift of 3.601, based on the comparison with the photometric redshift estimate \citep{boogaard19}. At this redshift, the 3-mm band also covers the [CI] 1-0 emission line. We extracted a spectral profile around the expected frequency of this line, however no line detection is found down to an rms of 0.26 mJy beam$^{-1}$ per 21 km s$^{-1}$ channel or 0.09 mJy beam$^{-1}$ per 200 km s$^{-1}$ channel. This places a limit to the line luminosity, assuming the [CI] line would have the same width than CO(4-3), of $L'_{\rm [CI]}=2.7\times10^9$ K km s$^{-1}$ pc$^2$ ($3\sigma$). Following \citet{bothwell17}, we compute an upper limit to the molecular gas mass from this [CI] line measurement \citep[see also][]{papadopoulos04, wagg06} using:

\begin{equation}
M({\rm H}_2)^{\rm CI} = 1375.8 \left( \frac{D_{\rm L}^2}{(1+z)} \right) \left( \frac{X_{\rm [CI]}}{10^{-5}} \right)^{-1}  \left( \frac{A_{\rm 10}}{10^{-7}} \right)^{-1} Q_{10}^{-1} F_{\rm [CI]},
\end{equation}

where $D_{\rm L}$ is the luminosity distance in Mpc, $X_{\rm [CI]}$ is the [CI]/H$_2$ abundance ratio, which we assume to be $3\times10^5$, and $A_{\rm 10}$ is the Einstein A coefficient equals to $7.93\times10^{-8}\ s^{-1}$. $Q_{10}$ is the excitation factor which we set at 0.6 and $F_{\rm [CI]}$ is the [CI] line intensity in units of Jy km s$^{-1}$. Thus, we find a $3\sigma$ limit for the [CI]-based molecular gas mass $M({\rm H}_2)^{\rm CI}<1.9\times10^{10}$ $M_\odot$. This limit is consistent with the molecular gas mass estimate derived from CO of $1.3\times10^{10}$ M$_\odot$. Note that this estimate extrapolates the CO(4-3) line emission down to CO(1-0) using a template obtained for massive BzK galaxies at $z=1.5$. If the CO SLED is steeper, with CO(4-3) and CO(1-0) closer to thermal equilibrium, the CO-derived gas mass would be in better agreement.  

\newpage

\section{Flux measurements in tapered cubes}
\label{sect:app2}
\begin{figure*}
\centering
\includegraphics[scale=0.5]{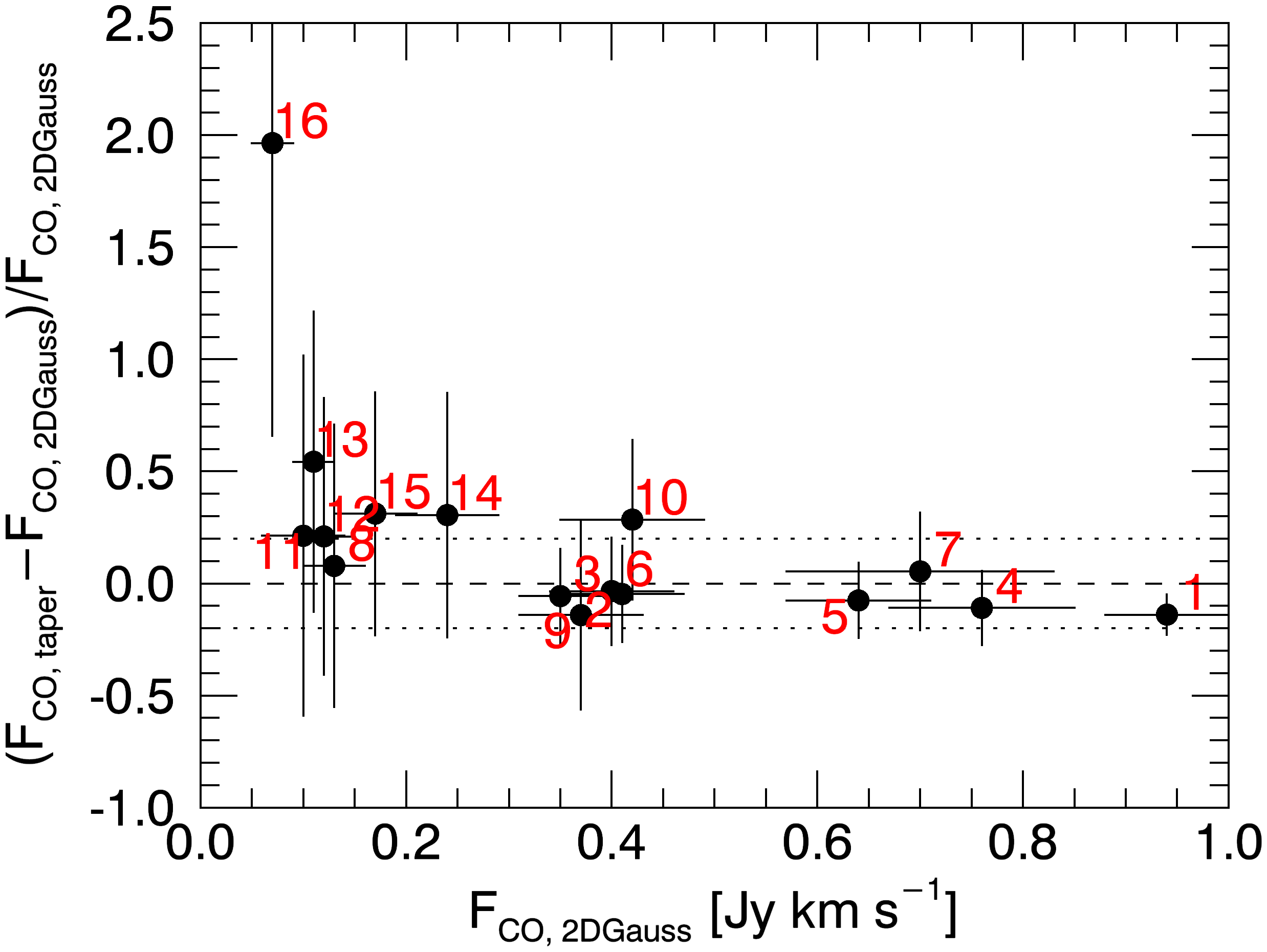}
\includegraphics[scale=0.5]{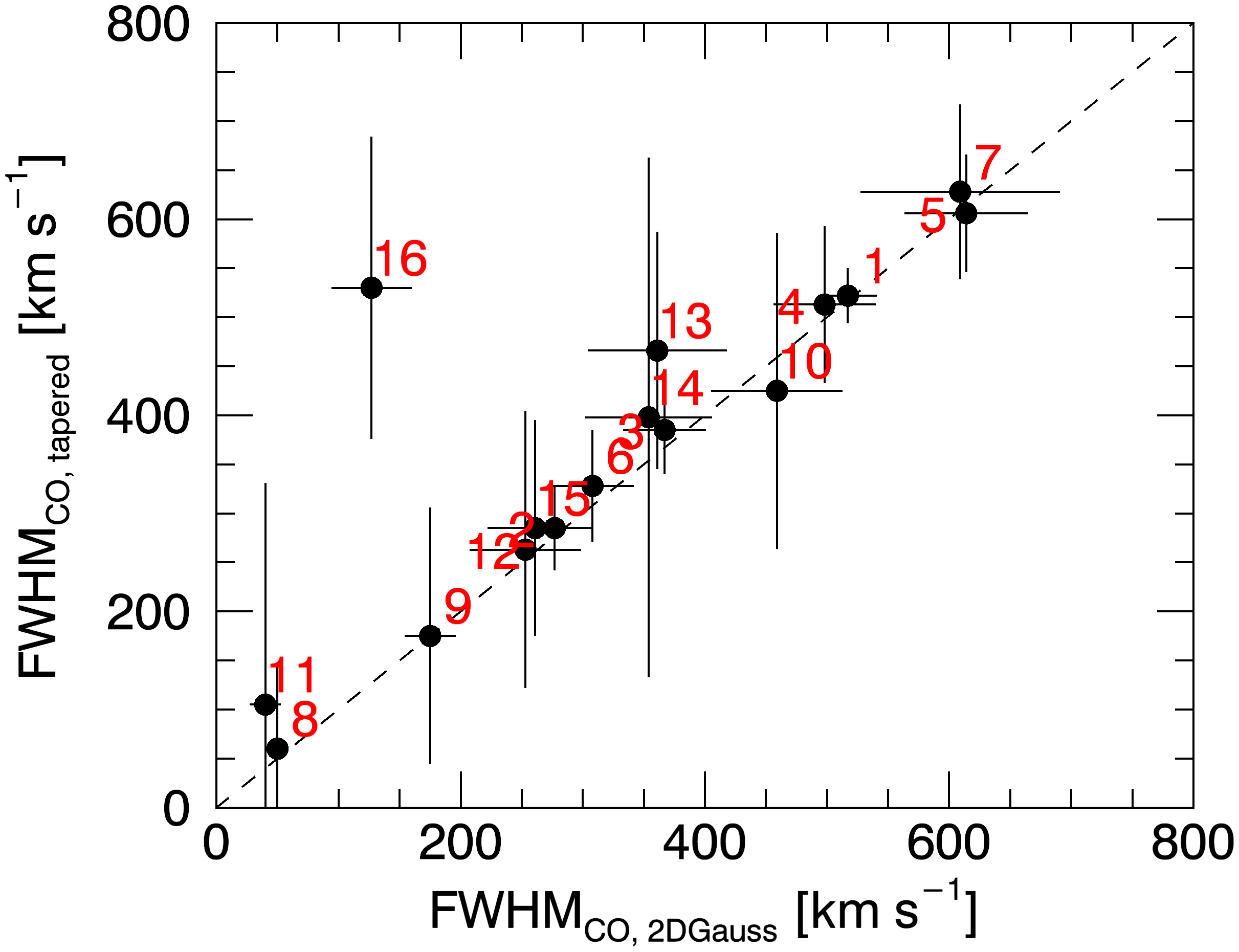}
\caption{Comparison of the CO flux measurements and line widths obtained from two-dimensional Gaussian fitting in the original resolution moment-0 maps versus the ones obtained from the measurements in the 3'' tapered cubes. Dotted lines in the top panel indicate lines of $20\%$ difference between these estimates. The dashed lines indicate the location of  identical estimates by both methods. \label{fig:comparison}}
\end{figure*}

We explored the possibility that we could be missing some flux due to sources being extended spatially. For this, we created a new version of the ASPECS band-3 data cube, tapered to an angular resolution of $\sim3''$, which should contain all of the extended CO line emission. We collapsed this cube, created the moment$-0$ images and computed the integrated fluxes as the value of the peak pixels in these images. Figure \ref{fig:comparison} shows the comparison of CO fluxes and line widths between these two estimates. The flux estimates for almost all sources are in excellent agreement within the uncertainties. 

In the case of source 3mm.16, the measurement in the tapered cube not only doubles the flux in the original one, but also yields a much larger line FWHM. This suggests significant extended low surface brightness emission, undetected in the original cube. Manual inspection of the cube, however, shows that the extra emission can be attributed to noise at large velocities ($>200$ km s$^{-1}$). All the other sources, however, have an excellent agreement between their measured FWHM. We thus use the original flux estimates throughout this paper.

\section{CO and optical sizes}

\label{sect:app3}
We used the CO moment-0 maps at original resolution (no tapering) to measure CO emitting sizes of the ASPECS sources. Here, we only focus on the 16 brighter CO-selected sources.  We used the {\sc CASA} task \texttt{imfit} to fit two dimensional Gaussian profiles to these images centered at the CO source positions. Due to the limited angular resolution and sensitivity of our observations, we did not attempt to fit more complicated profiles, which require more free parameters (i.e. Sersic profile). From this, we extracted the deconvolved semi-major and semi-minor axes of the fitted Gaussian profile ($B_{\rm maj}$, $B_{\rm min}$), and computed the half light radius $r_{\rm 1/2}$ by averaging these two (weighted by uncertainties). We computed the ellipticity of the profile as $e=B_{\rm maj}/B_{\rm min}$. We consider that the source is resolved in CO emission if either $B_{\rm maj}$ or $B_{\rm min}$ are measured at a significance above 3. The derived parameters are listed in Table \ref{table:4}.

\begin{table*}
\centering
\caption{Sizes of the ASPECS CO galaxies.$^{\dagger}$\label{table:4}}
\begin{tabular}{ccccccccc}
\hline\hline
ID & $z_{\rm CO}$ & $r_{\rm 1/2, CO}$ & $e_{\rm CO}$ & $r_{\rm 1/2, opt}$ & $n_{\rm opt}$ & $e_{\rm opt}$  \\
3mm. &  &  & (kpc)   &  & (kpc) &   &   \\
(1) &  (2) & (3) & (4) & (5) & (6) \\
\hline
 1 & 2.543 & $ 4.3\pm  0.4$ & $ 3.7\pm  3.4$ & $   1.7$ & $ 0.8$ & $ 0.8$ \\
 2 & 1.317 &  $\ldots$ & $\ldots$ & $   4.0$ & $ 2.2$ & $ 0.6$ \\
 3 & 2.453 &  $ 3.9\pm  0.5$ & $ 2.2\pm  0.7$ & $   5.1$ & $ 0.2$ & $ 0.3$ \\
 4 & 1.414 &  $ 5.0\pm  0.7$ & $ 3.7\pm  2.2$ & $   7.4$ & $ 0.5$ & $ 0.2$ \\
 5 & 1.550 &  $ 4.2\pm  0.7$ & $ 2.4\pm  1.1$ & $   8.3$ & $ 3.0$ & $ 0.4$ \\
 6 & 1.095 &  $ 4.5\pm  1.1$ & $ 1.2\pm  0.6$ & $   5.4$ & $ 1.1$ & $ 0.8$ \\
 7 & 2.697 & $\ldots$ & $\ldots$ & $   4.8$ & $ 0.9$ & $ 0.5$ \\
 8$^\ddagger$ & 1.382 &  $ 5.3\pm  1.6$ & $ 2.1\pm  1.4$ & $\ldots$ &  $\ldots$&  $\ldots$\\ 
 9 & 2.698 &   $\ldots$ & $\ldots$ & $   0.6$ & $ 7.2$ & $ 0.7$ \\
10 & 1.037 & $ 3.6\pm  1.1$ & $ 2.2\pm  1.8$ & $   2.5$ & $ 0.9$ & $ 0.5$ \\
11 & 1.096 & $ 3.6\pm  1.3$ & $ 4.1\pm  3.1$ & $   1.8$ & $ 0.2$ & $ 0.6$ \\
12$^\ddagger$ & 2.574 & $\ldots$ & $\ldots$ & $\ldots$ & $\ldots$ & $\ldots$ \\ 
13 & 3.601 & $ 4.0\pm  1.2$ & $ 1.8\pm  1.5$ & $   0.9$ & $ 2.1$ & $ 0.4$ \\
14$^X$ & 1.098 & $\ldots$ & $\ldots$ & $ \ldots$ & $\ldots$ & $ \ldots$ \\
15 & 1.096 &  $\ldots$ &$\ldots$ & $   6.0$ & $ 0.4$ & $ 0.4$ \\
16 & 1.294 &  $ 6.1\pm  1.7$ & $ 1.0\pm  0.6$ & $   4.8$ & $ 1.0$ & $ 0.5$ \\
\hline\hline
\end{tabular}
\\
\flushleft \noindent {\bf Notes.} $^{\dagger}$ For sources that were unresolved in CO emission, no sizes are provided. $^\ddagger$ Sources 3mm.8 and 3mm.12 do not have reliable optical counterparts and thus their optical sizes are not listed. $^X$ Source 3mm.14 does not have a reliable optical morphology estimate in the catalog of \citet{vanderwel12}. Columns: (1) Source ID. (2) CO redshift. (2) Half-light radius of the CO emission, assuming a Gaussian distribution (Sersic index of 0.5). (3) Ellipticity of the CO distribution. (4) Half-light radius of the optical emission, using a Sersic profile with index $n_{\rm opt}$ \citep{vanderwel12}. (5) Sersic index of the optical emission. (6) Ellipticity of the Sersic profile.
\end{table*}

In addition to the CO sizes, we use the structural parameters derived by \citet{vanderwel12} from the {\it HST} near-IR images of the CANDELS field. We remove sources 3mm.8 and 3mm.12 since their optical counterparts are contaminated by foreground structures. The parameters are listed in Table \ref{table:4}. \citet{vanderwel12} use Sersic profiles to fit these images, given the high resolution and signal of the rest-frame optical sources. Note that a two dimensional Gaussian profile is equivalent to a Sersic profile with index $n=0.5$. In some cases, the fitted profiles show Sersic index $n$ values above 2, indicative of a highly concentrated central source (for example, a bright central bulge, or an galactic nuclei). Thus, in these cases the derived values of the half light radius, $r_{\rm 1/2}$, in the {\it HST} images might not be necessarily comparable to the values derived for CO.

\begin{figure*}
\centering
\includegraphics[scale=0.6]{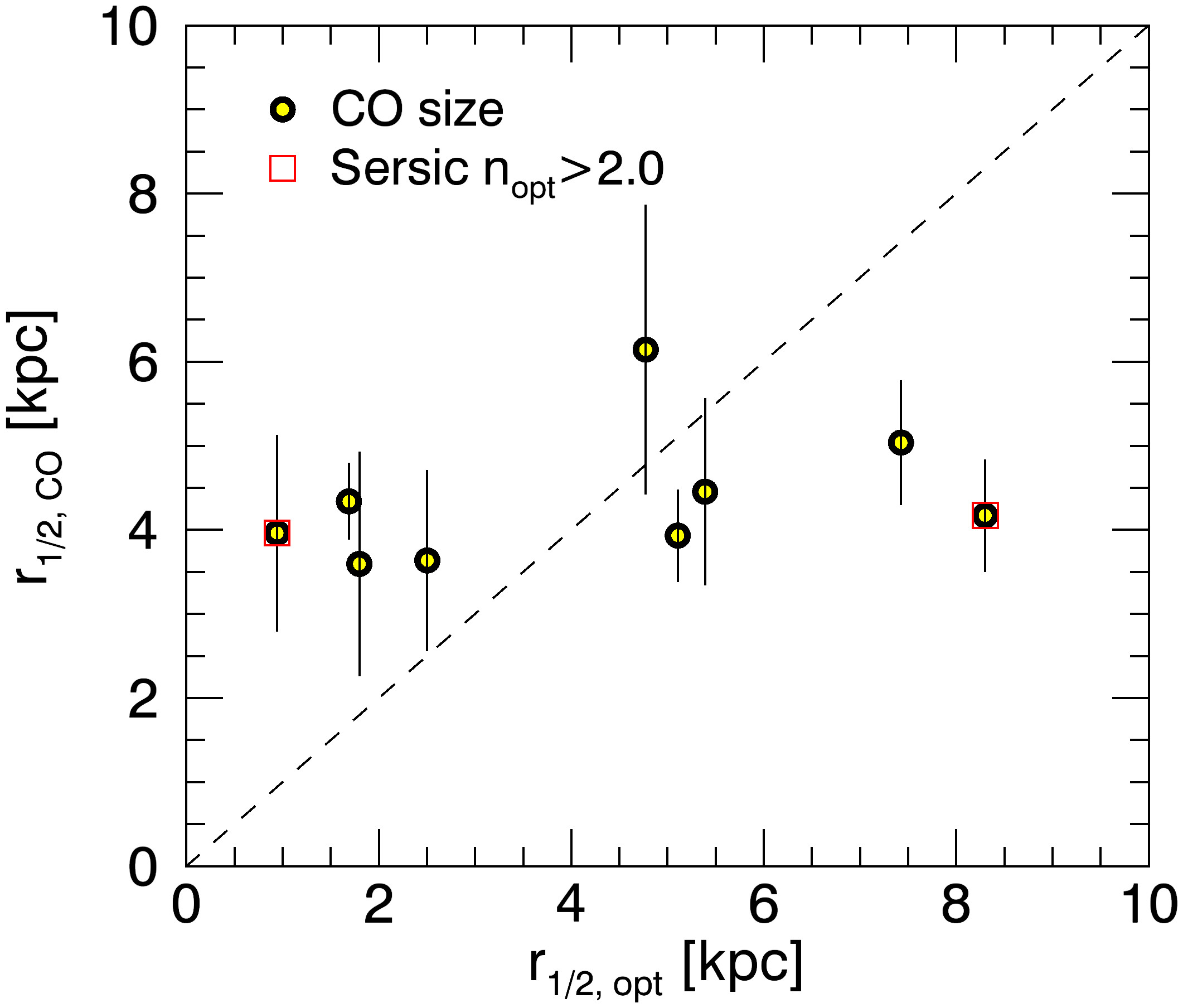}
\caption{Comparison between the rest-frame optical sizes derived by \citet{vanderwel12}  and the CO sizes measured from the ASPECS data. Only resolved sources in CO emission are considered. Red squares highlight sources for which the optical morphology indicates large Sercic indexes, which would indicate highly concentrated optical emission and thus might not be directly comparable to the CO estimates. \label{fig:sizes}}
\end{figure*}

\begin{figure*}
\centering
\includegraphics[scale=0.9]{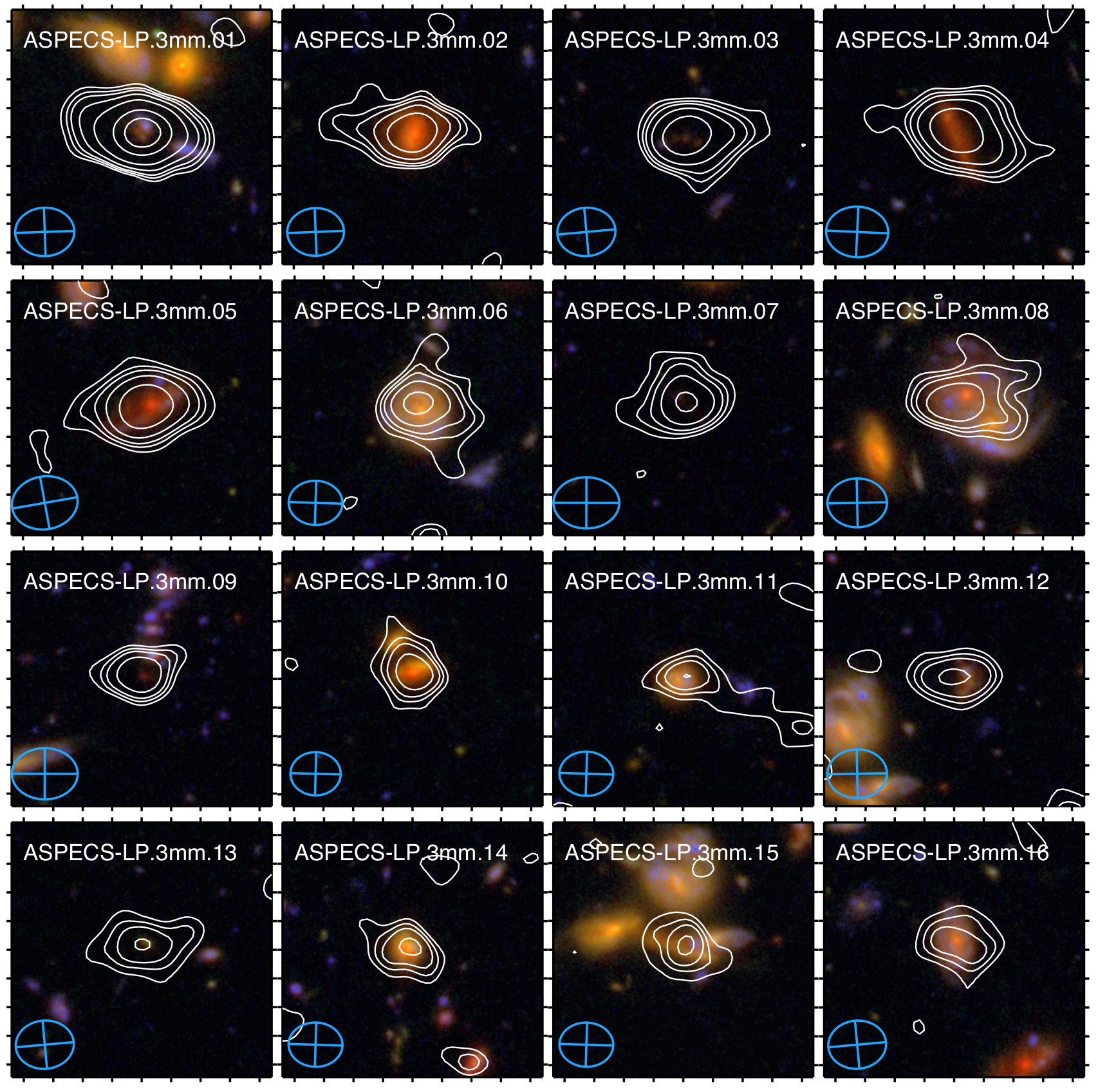}
\caption{Optical/near-IR postage stamps compared to the CO emission for the ASPECS CO-selected sample. {\it HST} RGB images (F435W, F850LP, F105W) are shown in the background with white contours overlaid representing the CO line emission at significances $2, 3, 4, 6, 10, 20$ and $30\sigma$, where $\sigma$ is the rms noise level of each CO moment-0 image. \label{fig:postage1}}
\end{figure*}

Figure \ref{fig:sizes} compares the CO and optical sizes ($r_{\rm 1/2}$) derived in this way. Figure \ref{fig:postage1} show a visual comparison of the optical/near-IR with the CO line emission morphologies. We find no clear correlation between the CO and the optical sizes. The CO sizes seem to stay relatively constant around $\sim4-6$ kpc, whereas the optical sizes span a significant range from $\sim$1.0 to 8.5 kpc. We note that even in cases where the CO emission is significantly resolved, as in sources 3mm.1, 3mm.3 or 3mm.5, the optical sizes show evident differences compared to the CO sizes. This suggests that (at least for these sources) the differences in size between CO and optical are physical, and not necessarily driven by the angular resolution and sensitivity limits of our data. This also may suggest that the CO sizes are relatively homogeneous in our sample. However, this result is limited by the fact that about half of our sample is currently unresolved. 

\section{CO kinematics}

\begin{figure*}[h] \centering \includegraphics[scale=0.9]{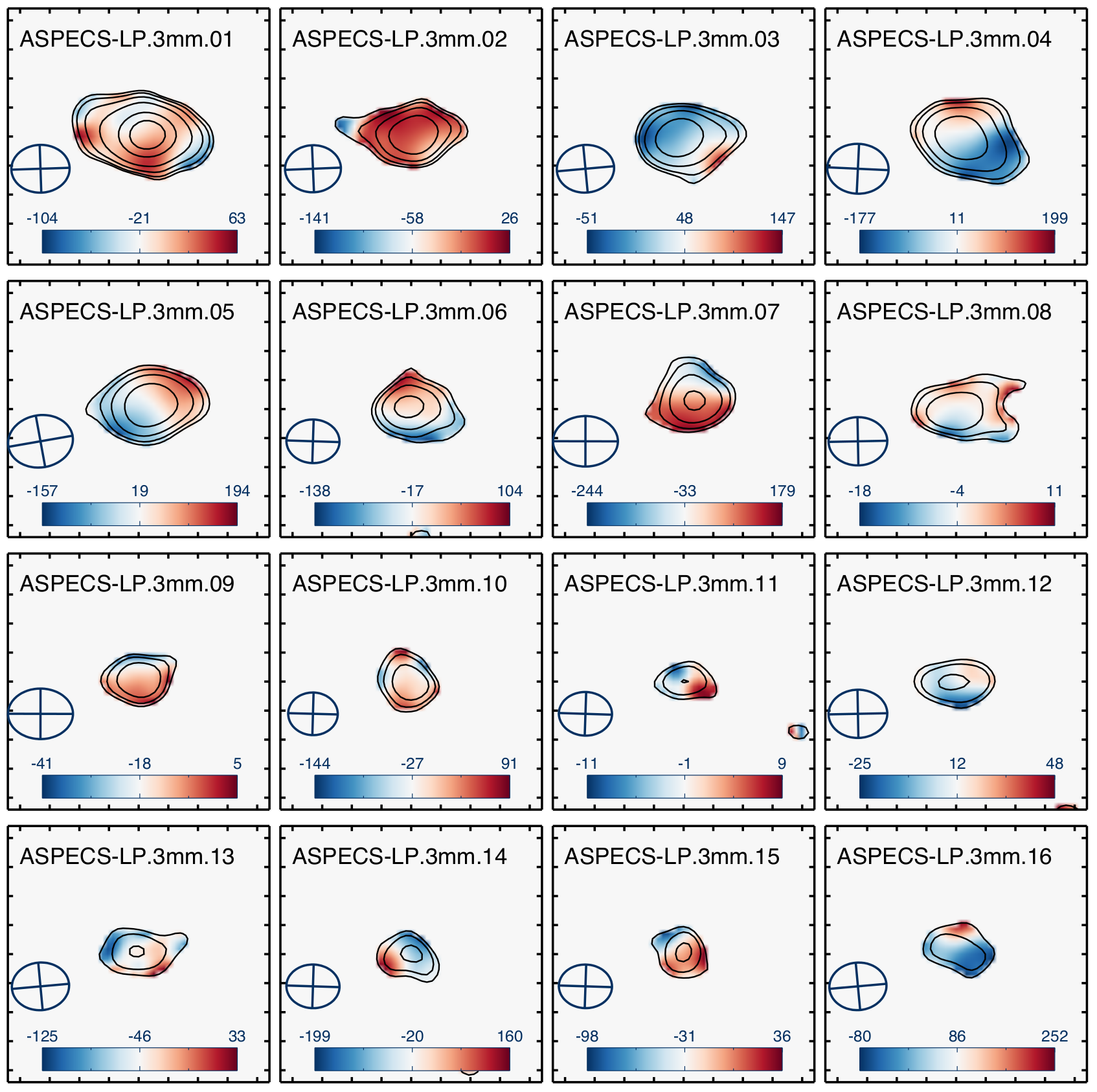}
\caption{Postage stamps of the CO moment-1 (velocity fields) toward the ASPECS CO sources. The background image represents the velocity field, with bluer and redder colors representing the approaching and receding CO components. The black contours show the moment-0 map, shown at levels 3, 6, 10, 15, 20 and 30$\sigma$, where $\sigma$ is the rms measured in this map. The color bar presented at the bottom of each panel shows the velocity scale in each case with respect to the CO central velocity, in units of km s$^{-1}$. The blue ellipse to the left side of each panel represents the beam size at the observed frequency.\label{fig:app2-mom1} } \end{figure*}

Since some of our galaxies were resolved in CO line emission, we computed CO moment-1 maps or velocity fields (see Fig. \ref{fig:app2-mom1}). In some cases, we clearly see velocity gradients suggestive of ordered gas rotation (3mm.4, 3mm.5, 3mm.6 and 3mm.7). In the particular case of 3mm.7, the CO emission is marginally resolved in one axis only, but the velocity field shows the structure clearly. Other cases with hints of velocity gradients are limited by the significance and resolution. Conversely, other cases where the emission in significantly resolved, as in 3mm.1, do not show evidence of rotation and suggest a dispersion dominated object.

We take advantage of the software $^{\rm 3D}$Barolo \citep{diteodoro15} to perform a tilted-ring modeling of the gas velocity field. Because of the coarse resolution of our data, we fix the ring inclination and position angle based on the Sersic fits performed on all the sources in the field by \citet{vanderwel12}, so that only the centroid of the line emission, the gas velocity and velocity dispersion are free in the fit. 

The ASPECS LP 3\,mm data were obtained in a relatively compact array configuration, thus the majority of the sources are only marginally resolved, and a proper dynamical analysis is not  feasible. Only three sources show a significant velocity gradient in the CO emission, which allows us to put loose constraints on the dynamical mass. The dynamical mass is derived as: $M_{\rm dyn} = R v_{\rm rot}^2 G^{-1}$. At $R$=$R_{\rm opt}$, the dynamical masses inferred for the ASPECS sources 3mm.4, 3mm.5, and 3mm.7 are $(8.1\pm2.4)\times10^{10}$\,M$_\odot$, $(2.7\pm0.8)\times10^{11}$\,M$_\odot$ and  $(1.4\pm0.5)\times10^{11}$\,M$_\odot$, respectively. 

\label{lastpage}

\end{document}